\documentclass[oneside,a4paper,english]{article}
\usepackage[a4paper,left=1in,right=1in,top=1in,bottom=1in]{geometry}
\usepackage{graphicx}
\usepackage{amsmath,amsfonts}
\usepackage{algorithm2e}
\usepackage{booktabs}

\usepackage{multirow}

\usepackage{bm}
\graphicspath{{figs/}}

\title{An extended Elrod-Adams model to account for backpressure and blow-by inception}

\author{Alfredo Jaramillo$^a$, Gustavo C. Buscaglia$^a$\\
	{\small
		(a) Instituto de Ci\^encias Matem\'aticas e de Computa\c{c}\~ao,
		Universidade de S\~ao Paulo, 13560-970 S\~ao Carlos, Brazil
}}

\newcommand{\parder}[2]{\frac{\partial #1}{\partial #2}}
\newcommand{\tm}[1]{\textnormal{\scriptsize{#1}}}

\newcommand{\pcc}{p_{\textnormal{\scriptsize{cc}}}}

\begin{document}
	\vspace{3cm}
	
	\maketitle
	
	
	\begin{abstract}
	  The Piston-Ring-Liner system is the main tribological component of internal combustion engines. The Elrod-Adams  model is customarily used to numerically assess the hydrodynamics of different ring designs and liner surface treatments. However, that model does not incorporate the backpressure boundary condition, which in this case corresponds to the combustion chamber pressure and is quite significant. In this contribution a model is proposed that imposes the combustion-chamber pressure in a mass-conserving way, together with an effective algorithm for its numerical approximation. The new model incorporates the pressure difference across the ring, which is shown to have a substantial effect on the predicted friction force and MFT. The model is further elaborated so as to provide a criterion for predicting blow-by inception.
		
		\medskip

		
	\end{abstract}
	
	\section{Introduction}\label{sec:introduction}
	
	Current hydrodynamical models allow to estimate load carrying capacity, friction losses and minimum film thickness (MFT) of tribological devices that involve challenging physical phenomena such as cavitation, which is a non-linear rupture of the fluid continuity \cite{Dowson1979}. Regarding the Piston-Ring-Liner (PRL) mechanism, many studies have addressed its tribological performance setting the Combustion Chamber Pressure (CCP) as a boundary condition along the Reynolds cavitation model for pressure \cite{Jeng1992,Han1998,Mufti2008,Gadeshi2012,Mezghani2012,Meng2014,Morris2014,Medina2015, Kligerman2015, Liu2016,Usman2016,Fang2017}. For instance, Kligerman et al \cite{Kligerman2015} compared the predicted friction force and MFT between two models: The ``Real engine simulation'' where the CCP was considered as a boundary condition, and the ``Approximate solution'', based on a quasi-static simulation where a boundary condition equal to the ambient pressure was imposed. The authors found that the MFT was strongly affected by the CCP time variation, and that the ``Approximate solution'' underestimated up to 30\% the time-averaged friction force. 
	
	However, mass-conservation is a mandatory property of numerical schemes when considering, as is the current trend, the microscopic features of liner surfaces \cite{Ausas2007,Qiu2009}. Recently, many studies on the PRL performance have adopted the mass-conserving Elrod-Adams model \cite{Chong2011,Tomanik2013,Checo2014a,Checo2014b,Profito2016,Profito2017,Checo2017,Hu2018}. Unfortunately, that model assumes the boundary condition for pressure as being equal to the cavitation pressure, which is invalid for the PRL mechanism since the CCP can assume values in the order of 100 atm.
	
	In this work, an extension of the Elrod-Adams model for non-homogeneous boundary conditions is proposed based on the numerical model presented in \cite{Jaramillo2016a,Jaramillo2016b}. Friction and MFT for a textured ring are computed with the extended model and a numerical comparison with both the Elrod-Adams and the Reynolds models is performed.
	
	A particular characteristic of the PRL operating conditions is that gas leakage can take place through the gaps between piston and ring and/or between ring and liner. This undesirable effect is known as blow-by and is associated to power-loss and air pollution \cite{Cheng2015}.  Performing two-dimensional simulations it is shown here that the extended model can be used to predict blow-by inception, which manifests itself as the development of \emph{channels} of gas/vapor through the gap between the ring and the liner.
	
	This paper's structure is as follows: In Section \ref{sec:math-model} the extended Elrod-Adams model is described in terms of a restriction on the pressure gradient at the rupture boundary, and typical models for the PRL geometry and the ring dynamics are introduced. The numerical method to solve the extended model coupled to the ring dynamics is presented in Section \ref{sec:num-methods}. A comparison of the resulting tribological performance between  the extended model and both the Elrod-Adams and the Reynolds model is performed in Section \ref{sec:num-simulations}. In that section we also expose predictions of an encounter between the reformation and the rupture boundaries for rings of different degree of conformity, which is put forward as a criterion for blow-by. Concluding remarks are drawn in the final section.
	
	\section{Mathematical models}\label{sec:math-model}
\begin{figure}[h]\centering
		\includegraphics[width=0.8\linewidth]{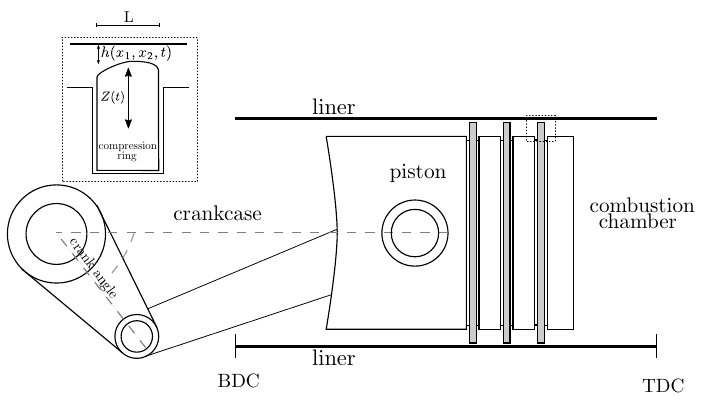}
		\caption{Scheme of the Piston-Ring-Liner system.}\label{fig:sec2:PRL}
	\end{figure}

	The main components of the PRL system are depicted in Fig. \ref{fig:sec2:PRL}, here a simplified modeling of that system is assumed as described in the next section. The fundamental scales are given in Table \ref{tab:scales} and the nomenclature for the involved quantities is given in Table \ref{tab:nomenclature}.
	\begin{table}[h!]
		\begin{center}
			\begin{tabular}{llll}
				\toprule 
				Symbol & Value & Units & Description \\ 
				\midrule
				$H$ & $1\times 10^{-6}$ & m & Gap thickness\\
				$L$ & $1\times 10^{-3}$ & m & Scale on the $x_1$-$x_2$ plane\\
				$\mu$ & $4\times 10^{-3}$ & Pa$\cdot$s & Lubricant viscosity\\
				$U$ & $10$ & m/s & Reference speed along the $x_1$ axis \\
				\bottomrule
			\end{tabular} 
			\caption{Fundamental scales.}\label{tab:scales}
		\end{center}
	\end{table}
	\begin{table}[h!]
		\begin{center}
			\begin{tabular}{lll}
				\toprule
				Quantity&Description& Scale\\
				\midrule
				$x_1$ & Longitudinal coordinate & $L$\\
				$x_2$ & Transverse coordinate & $L$\\
				$t$ & Time & $L/U$\\
				$h,h_\tm{ring},h^\delta_\tm{wear}$ & Functions defining the surfaces gap & $H$\\
				$h_\tm{feed}$ & Film thickness at the boundaries & $H$\\
				$u$ & {Liner's} speed along $x_1$ & $U$\\
				$Z$ & Minimum gap in time & $H$\\
				{$L_\tm{ring}$} &  {Ring's length along $x_1$} & {$L$}\\
				$B_\tm{r}$ &  Bore radius & $L$\\
				$p$ & Hydrodynamic pressure & $6\mu\,UL/H^2$\\
				$W^\tm{h},W^\tm{a},W^\tm{c}$ &  Radial forces per unit width & $6\mu\,UL^2/H^2$\\
				$m$ & Ring's mass per unit width & $6\mu L^4/(H^3U)$ \\
				$\mu_\tm{c}$ & Boundary friction coefficient ($=0.11$) & - \\
				\bottomrule
			\end{tabular} 
			\caption{Nomenclature.}\label{tab:nomenclature}
		\end{center}
	\end{table}

	\subsection{Geometrical model and ring dynamics}
	\label{sec:model-geo-adim}
	
\begin{figure}[h]\centering
		\includegraphics[width=0.6\linewidth]{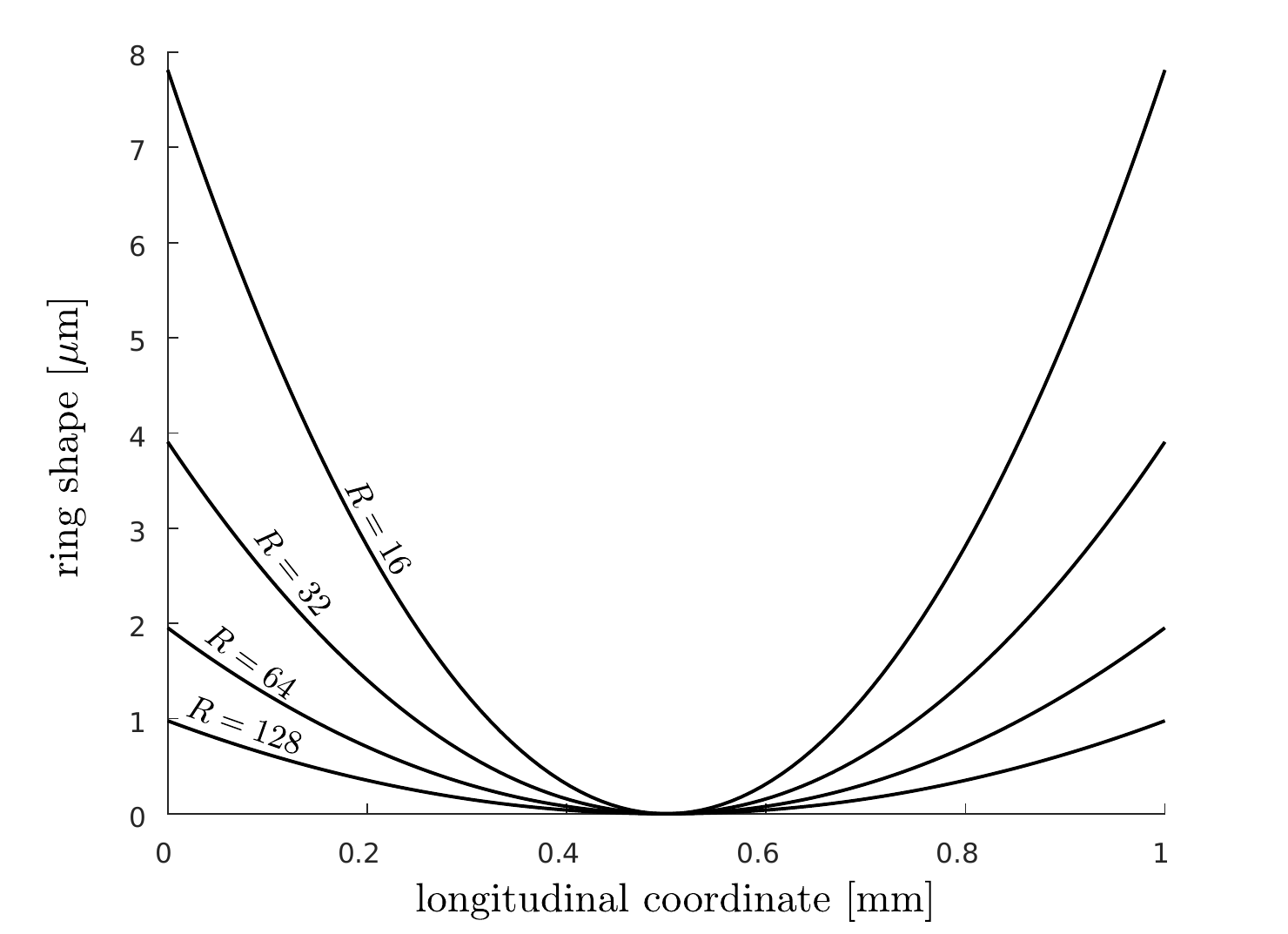}
		\caption{Ring shape along the $x_1$-axis for different values of $R$.}\label{fig:sec2:ring-shape}
	\end{figure}

Both surfaces (ring and liner) are considered to be smooth (no surface roughness) for simplicity. The gap between the surfaces is modeled by the function
		\begin{equation}
	h(x_1,x_2,t)=Z(t)+h^R_\tm{ring}(x_1)+h^\delta_\tm{wear}(x_2),
	\label{eq:sec2:h-gap}
	\end{equation}
	where $Z$ is the dynamic variable accounting for the expansion/contraction of the ring, $h_\tm{ring}$ is the function describing the ring's shape, defined as satisfying $\min_{x\in [0,1]} h^R_\tm{ring}(x_1)=0$, depicted in Fig. \ref{fig:sec2:ring-shape} and written as
	\begin{equation}
	h_\tm{ring}^R(x_1,x_2)=\frac{L}{H}\frac{\left(x-0.5\right)^2}{2R},
		\label{eq:hring}
	\end{equation}
	and $h^\delta_\tm{wear}$ is a function modeling a non-homogeneous wear of the surfaces along the circumferential direction, also satisfying $\min_{x\in [0,2\pi B_\tm{r}]} h^\delta_\tm{wear}(x_2)=0$ and assumed for simplicity to have a Gaussian profile centered at the circumferential angle equal to $\pi$:
	\begin{equation}
	h_\tm{wear}^\delta(x_2)=\delta \exp\left(-(x_2-\pi B_\tm{r})^2/c^2\right)\qquad x_2\in[0,2\pi B_\tm{r}].	\label{eq:sec2:h-wear}
	\end{equation}
	
	The ring follows a reciprocal motion along the $x_1$-axis and its speed is denoted by $u(t)$. The dynamics of $Z(t)$ is determined by Newton's equation:
	\begin{equation}
	m \frac{d^2Z}{dt^2}= W^\textnormal{h}(t)+W^\textnormal{ps}+W^\textnormal{con}(h){+W^\tm{cc}(t)}.	\label{eq:sec2:newton-Z}
	\end{equation}
	The first term on the left-hand side of this equation corresponds to the hydrodynamical force per unit width (in the circumferential direction), computed as
	\begin{equation}
	W^\tm{h}(t)=\frac{1}{B_r}\int_0^{2\pi B_\tm{r}}\int_a^b p(x_1,x_2,t)\,dx_1\,dx_2,\label{eq:sec2:rforce-hyd}
	\end{equation}
	the second and third terms correspond to the applied load (datum) and the contact load, respectively, the latter being computed by means of the Greenwood-Tripp model \cite{Greenwood1970} as
		\begin{equation}
	W^\tm{con}(h)=\frac{1}{B_r}\int_0^{2\pi B_\tm{r}}\int_a^b p^\tm{con}(h(x_1,x_2,t))\,dx_1\,dx_2,\label{eq:sec2:rforce-pcon}
	\end{equation}
	with the contact pressure $p^\tm{con}$ given by
	\begin{equation}
		p^\tm{con}(h)=\pi\frac{16\sqrt{2}}{15}\left(\eta \beta \sigma\right)^2\sqrt{\sigma/\beta}E'F_{5/2}\left(h/\sigma\right),
	\end{equation}
	where $E'$ is the composite modulus of elasticity, and $\eta$, $\beta$ and $\sigma$ are the asperity density, radius of curvature and composite standard deviation of the asperities height. The value of $F_{5/2}\left(h/\sigma\right)$ corresponds to an integral that is approximated as done by Panayi-Schock in \cite{Panayi2008}. {The last term in Eq. \eqref{eq:sec2:newton-Z} corresponds to the pressure exerted on the back of the ring, pushing it towards the liner and modeled by
\begin{equation}
	W^\tm{cc}(t)=-\gamma\, \pcc(t)\,L_\tm{ring}
\end{equation}	
where $\gamma$ is the Back-pressure factor and $L_\tm{ring}$ is the ring's length along the $x_1$ direction.
}
	
		
	The friction force exerted by the fluid on the ring at time $t$ along the $x_1$-axis is given by \cite{Checophd}
	\begin{equation}
	F(t)=\frac{1}{B_r}\int_0^{2\pi B_\tm{r}}\int_a^b \left(\frac{\mu u }{h}g(\theta)-\frac{h}{2}\parder{p}{x_1}-p\parder{h_\tm{ring}}{x_1}+\mu_\tm{c}\,p^\tm{con}\right)\,dx_1\,dx_2,	\label{eq:sec2:friction}
	\end{equation}
	where
	\begin{equation}
		g(\theta)=\begin{cases}
		\theta & \mbox{if }\theta>\theta_\tm{s}\\
		0 & \mbox{otherwise}
		\end{cases},
	\end{equation}
	and $\mu_\tm{c}$ is the boundary friction coefficient. The threshold $\theta_\tm{s}$ is interpreted as the minimum oil fraction for shear forces to be transmitted between the surfaces and it was set to 0.95 (e.g., \cite{Checo2014a}).
		
	\subsection{Hydrodynamics and cavitation modeling}
	\label{sec:hydrodynamics-cavitation}
\begin{figure}[ht!]\centering
		\includegraphics[width=1\linewidth]{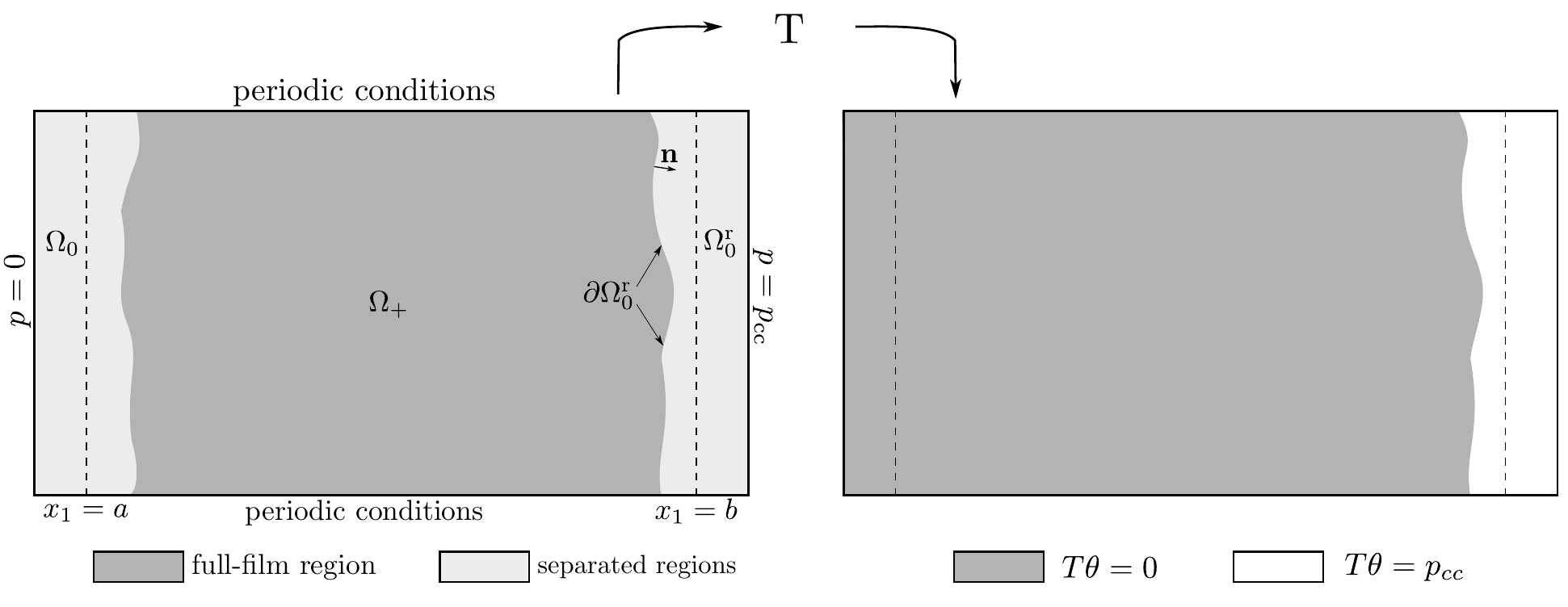}
		\caption{Scheme of the operator $T$.}\label{fig:sec2:operator_T}
	\end{figure}

	The hydrodynamical pressure $p$ and the saturation field $\theta$ are modeled by an extension of the Elrod-Adams cavitation model that reads
	\begin{equation}
	\nabla\cdot \left(\frac{h^3}{2}\nabla p\right)=\frac{u(t)}{2}\parder{h\theta}{x_1}+\parder{h\theta}{t}	,\label{eq:sec2:reynolds}
	\end{equation}
	along the restrictions for the unknown fields:
	\begin{equation}
		p\geq T(\theta)\qquad\mbox{and}\qquad 0\leq \theta \leq 1
	\end{equation}
	the boundary conditions:
	\begin{equation}
	\begin{cases}
	p(0,x_2)=0,&p(1,x_2)=\pcc(t)\\
	\theta(0,x_2)=\frac{h_\tm{feed}}{h(0,x_2)},&\theta(1,x_2)=\frac{h_\tm{feed}}{h(1,x_2)} 
	\end{cases}\qquad \mbox{for } x_2\in [0,2\pi B_\tm{r}]
	\end{equation}
	and the complementarity conditions:
	\begin{align}
		\left(p-T(\theta)\right)\left(1-\theta\right)&=0.
	\end{align}
	The novelty of this model lies in the operator $T$. When $T(\theta)\equiv 0$ one recovers the classical Elrod-Adams model, which essentially assumes a zero combustion-chamber pressure. Let us denote $\Omega_0=\{\mathbf{x}\in\Omega:\theta(\mathbf{x})<1\}$, the region of the domain where the film is not complete, which in general has many connected components. The basic idea of the model proposed here is that the connected component of $\Omega_0$ that touches the right boundary must have $T$ equal to the combustion-chamber pressure, while in the others $T\equiv 0$. More specifically, denoting by $\Omega_0^\tm{r}$ the connected component of $\Omega_0$ that intersects the set $\{(x_1,x_2)\in\Omega:x_1=1\}$, we define
	\begin{equation}
	T(\theta)(\mathbf{x})=\begin{cases}
	\pcc(t) & \mbox{for }\mathbf{x}\in\Omega_0^\tm{r}\\
	0 & \mbox{otherwise}
	\end{cases}.	\label{eq:sec2:operator-T}
	\end{equation}
	An example of this is shown in Fig. \ref{fig:sec2:operator_T}. Notice that if the right and left sides of the system are connected by a ``channel'' of imcomplete film, i.e., if the intersection $\Omega_0\cap\Omega_0^\tm{r}$ is non-empty and $\pcc(t)>0$, the model becomes undefined. This condition intuitively determines the blow-by of combustion chamber gases, and will be so considered hereafter.

	As in the case of the standard Elrod-Adams model, the extended model consists of a transport equation on $\Omega_0$ and an elliptic equation on the full-film region. Though the standard model (interpreted in the distribution sense) enforces mass conservation throughout the domain, when $\pcc>0$ a supplementary boundary condition must be supplied at rupture points. In fact, on the interface $\partial \Omega_+$ the mass conservation condition can be written, assuming $h$ to be continuous, as
	\begin{equation}
		\parder{p}{\mathbf{n}}=2\frac{(1-\theta_-)}{h^2}\left(\frac{U}{2}\mathbf{e}_\tm{c}-\mathbf{V}^\tm{i}\right)\cdot \mathbf{n},\label{eq:sec2:rankine}
	\end{equation}
	where $\theta_-$ is the value of $\theta$ at the cavitated/separated side, $\mathbf{e}_\tm{c}$ is the unit vector along the Couette flux, $\mathbf{V}^\tm{i}$ is the interface velocity and $\mathbf{n}$ is the outward normal to $\partial \Omega_+$. This last condition implies
	$$\begin{cases}
	\parder{p}{\mathbf{n}}\geq 0&\mbox{at rupture points}\\
	\parder{p}{\mathbf{n}}\leq 0&\mbox{at reformation points}
	\end{cases},$$
	which when $T\equiv 0$ corresponds to the null-gradient condition at the rupture points but, when $\pcc > 0$, is not enough to assure the well-posedness of the extended model \cite{Jaramillophd,Buscaglia2013,Ausas2013}. To tackle this issue, in this work the condition $\parder{p}{\mathbf{n}}= 0$ is imposed at the rupture points. This can be summarized as
	\begin{equation}
	\parder{p}{\mathbf{n}}\leq 0\label{eq:sec2:non-positive-gradient}\qquad \mbox{in }\partial\Omega_+.
	\end{equation}
	
	\section{Numerical Method}
	\label{sec:num-methods}

The domain $\Omega$ is divided in $N_{x_1}\times N_{x_2}$ cells and a finite volume scheme for Eq. \eqref{eq:sec2:reynolds} is used. For this, the $x_1$ flux component going from node $(i-1,j)$ to node $(i,j)$ is discretized by
\begin{equation}
-\frac{h^3}{2}\parder{p}{x_1}+\frac{u}{2}\,h\theta \simeq -\frac{1}{2}\frac{(h^n_{i-1,j})^3+(h^n_{i,j})^3}{2}\frac{p^n_{i,j}-p^n_{i-1,j}}{\Delta x_1}+\frac{u}{2}\,h^n_{i-1,j}\theta^n_{i-1,j},
\label{eq:sec3:fvs-ea}
\end{equation}
where an upwind approximation is used for the Couette term. Balancing the fluxes on each cell and discretizing time along an implicit scheme for the temporal term $\parder{h\theta}{t}$ one gets the system of equations:
\begin{align}
a^{00}_{i,j}\,p^n_{i,j}+e^{00}_{i,j}\,\theta^n_{i,j} &= C_{i,j}(\bm{p}^n,\bm{\theta}^n),\label{eq:sec3:a_e_C_discr}\\
p_{i,j}^n&\geq T\left(\bm{\theta}^n\right)_{i,j},\label{eq:sec3:discr_restr_p}\\
0\leq \theta_{i,j}^n&\leq 1,\label{eq:sec3:discr_restr_theta}\\
\left(p_{i,j}^n-T\left(\bm{\theta}^n\right)_{i,j}\right)\left(1-\theta_{i,j}^n\right)&=0,\label{eq:sec3:compl_rest_discr}
\end{align}
where 
\begin{align}
C_{i,j}(\bm{p}^n,\bm{\theta}^n)=&-a_{i,j}^{-0} p^n_{i-1,j}-a_{i,j}^{+0} p_{i+1,j}^n
-a_{i,j}^{0+} p^n_{i,j+1}-a_{i,j}^{0-} p_{i,j-1}^n-e_{i,j}^{-0} \theta_{i-1,j}^n+f_{i,j}^n,
\end{align}
with
$$
\begin{array}{ll}
a_{i,j}^{00}=s_{i+1,j}^n+s_{i-1,j}^n+(\Delta x_1/\Delta x_2)^2\left(s_{i,j+1}^n+s_{i,j-1}^n\right), & e_{i,j}^{00}=\left(u\,\Delta x_1+2\Delta x_1^2 / \Delta t\right) h^n_{i,j},\\
a_{i,j}^{+0}=-s^n_{i+1,j}, & a_{i,j}^{-0}=-s^n_{i-1,j},\\
a_{i,j}^{0-}=-(\Delta x_1/\Delta x_2)^2\,s^n_{i,j-1}, & a_{i,j}^{0+}=-(\Delta x_1/\Delta x_2)^2\,s^n_{i,j+1}, \\
e_{i,j}^{-0}=-u\,\Delta x_1\, h^n_{i-1,j}, & s^n_{i\pm 1,j\pm 1}=\frac{1}{2}\left((h_{i,j}^n)^3+(h_{i\pm 1,j\pm 1 }^n)^3\right),\\
f_{i,j}^n = 2\Delta x_1^2/\Delta t\, h_{i,j}^{n-1}\theta _{i,j}^{n-1}.&
\end{array}$$

Notice that $a^{00}_{i,j},e^{00}_{i,j},f_{i,j}^n$ are non-negative, and each term $a^{+0}_{i,j}$, $a^{-0}_{i,j}$, $a^{0+}_{i,j}$, $a^{0-}_{i,j}$, $e_{i,j}^{-0}$ is non-positive, thus $C_{i,j}(\bm{p}^n,\bm{\theta}^n)\geq 0$.
This type of discrete system was already studied for the case $T\equiv 0$ (standard Elrod-Adams algorithm) in \cite{Alt1980} by means of a fixed-point method and in \cite{Marini1986} where the convergence of some Gauss-Seidel-like algorithms was proved.

Dropping the superscript $n$ and denoting by $N$ the number of unknowns, the system of equations \eqref{eq:sec3:a_e_C_discr}-\eqref{eq:sec3:compl_rest_discr} can be written as a fixed-point problem by defining the operator $B_T:\mathbb{R}^{N}\times \mathbb{R}^{N} \mapsto \mathbb{R}^{N}\times \mathbb{R}^{N}$ as
\begin{equation}
B_T(\bm{p},\bm{\theta})_{i,j}=\begin{cases}
\left(\frac{C_{i,j}(\bm{p},\bm{\theta})-e_{i,j}^{00}}{a_{i,j}^{00}\,(T\bm{\theta})_{i,j}},\,1\right) & \text{if}\,\,\,\, \frac{C_{i,j}(\bm{p},\bm{\theta})-e_{i,j}^{00}}{a_{i,j}^{00}\,(T\bm{\theta})_{i,j}}\geq  0\\
\left(T(\bm{\theta})_{i,j},\, \frac{C_{i,j}(\bm{p},\bm{\theta})-a_{i,j}^{00}\,T(\bm{\theta})_{i,j}}{e^{00}_{i,j}}\right) & \text{if}\,\,\,\, \frac{C_{i,j}(\bm{p},\bm{\theta})-e_{i,j}^{00}}{a_{i,j}^{00}\,(T\bm{\theta})_{i,j}}<
0
\end{cases},\label{eq:sec3:def-operator-BT}
\end{equation}
This way, Equation (\ref{eq:sec3:a_e_C_discr}) can be rewritten as the fixed point problem
\begin{equation}
B_T(\bm{p},\bm{\theta}) = (\bm{p},\bm{\theta}).\label{eq:sec3:fixed_point}
\end{equation}

Let us denote by $I=I_{0}\overset{\circ}{\cup} I_\textnormal{r}$ the set of cells indexes for which $\bm{p}$ and $\bm{\theta}$ are unknowns, where $I_\textnormal{r}$ is the subset of cells in the region $\Omega_0^\textnormal{r}$ and $I_{0}=I\setminus I_\textnormal{r}$. A solution to Eq. \eqref{eq:sec3:fixed_point} accomplishes automatically the conditions \eqref{eq:sec3:discr_restr_p} and \eqref{eq:sec3:compl_rest_discr} for any $(i,j)\in I$. However, the only immediate condition accomplished by $\theta_{i,j}$ is that $\theta_{i,j}\leq 1$. In fact, we have 
\begin{equation*}
C_{i,j}(\bm{p},\bm{\theta}) -a_{i,j}^{00}T(\bm{\theta})_{i,j} =f_{i,j}+S\Delta x_1 h_{i	-1,j} \theta_{i-1,j}+ \,\Delta''_{i,j},
\end{equation*}
where $f_{i,j}+S\Delta x_1 h_{i	-1,j} \theta_{i-1,j}\geq 0$ and
\begin{align*}
\Delta''_{i,j}&=s_{i-1,j}\left(p_{i-1,j}-T(\bm{\theta})_{i,j}\right)+s_{i+1,j}\left(p_{i+1,j}-T(\bm{\theta})_{i,j}\right)+\\
&+q^2\left\{s_{i,j-1}\left(p_{i,j-1}-T(\bm{\theta})_{i,j}\right)+s_{i,j+1}\left(p_{i,j+1}-T(\bm{\theta})_{i,j}\right)\right\}\nonumber,
\end{align*}
with $q$ a constant not depending on $\bm{p}$ nor $\bm{\theta}$. Thus, $\theta_{i,j}$ may be negative in the cells of $I_\textnormal{r}$ for which some of its neighbors belongs to the set $I_0$. This issue can be overcome by defining $T$ in such a way that $\Delta''_{i,j}\geq 0$.

\begin{figure}[h!]
	\centering
	\def\svgwidth{0.9\linewidth}
	\includegraphics[width=1\linewidth]{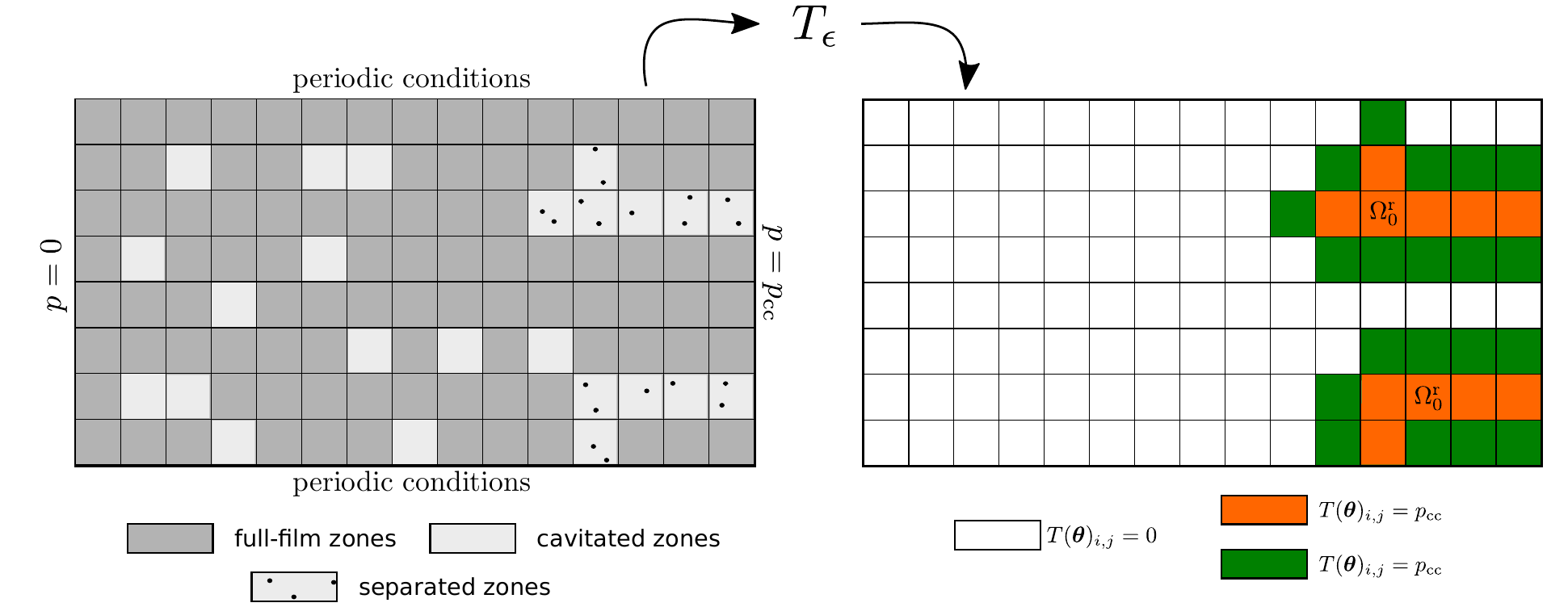}
	\caption{Illustration of the action of the discrete operator $T_\epsilon$.}
	\label{fig:sec3:operator-T-discr}
\end{figure}

The discrete version of the operator $T=T_\epsilon$ is thus defined as 
\begin{equation}
T_\epsilon(\bm{\theta})_{i,j}=
\begin{cases}
\pcc(t)  &\text{ if }(i,j)\in I_r \cup V^\epsilon \\
0 				& \text{ if } (i,j)\notin I_r \cup I^\epsilon
\end{cases},\label{eq:sec3:def-operator-T-discr}
\end{equation}
Where $I^\epsilon$, $\epsilon\in\{0,1\}$, corresponds to 
the indexes $\left(i,j\right)$ of the cells contained in $\Omega_+$ such that some of its neighbor cells $\{(i\pm 1,j),(i,j\pm 1)\})$ belong to $\Omega^\textnormal{r}_0$. { By extending the value of $T$ to the neighboring cells the zero normal derivative at rupture points is automatically imposed and, furthermore, the discrete values of $\theta$ become automatically non-negative. }
Figure \ref{fig:sec3:operator-T-discr} illustrates the action of the discrete operator $T_\epsilon$. In \cite{Alt1980} an algorithm to solve Eq. \eqref{eq:sec3:fixed_point} for the case $T_\epsilon\equiv 0$ is proposed. That numerical procedure is presented in Algorithm \ref{alg:sec3:alg-gs_reynolds}, which is also based on an adaptation of the numerical strategy proposed by Ausas et al \cite{Ausas2009} to solve the coupling of Eq. \eqref{eq:sec3:fixed_point} with a Newmark scheme for the Newton's equation \eqref{eq:sec2:newton-Z}. The computation of $T(\bm{\theta})$ required at each iteration is performed by a flooding algorithm that starts from the right boundary (the combustion chamber side).

\begin{algorithm}[h]
	\SetKwFunction{FRecurs}{FnRecursive}%
	\KwIn{$h$: gap function; $\bm{p}^{n-1}$, $\bm{\theta}^{n-1}$, $Z^{n-1}$, $V^{n-1}$: initial guess}
	\LinesNumbered
	\Begin{
		$k=1$\;
		$\bm{p}^{n,0}=\bm{p}^{n-1}$; $\bm{\theta}^{n,0}=\bm{\theta}^{n-1}$; $Z^{n,0}=Z^{n-1}$\;
		$\bm{p}^{n,k}=\bm{p}^{n,0}$; $\bm{\theta}^{n,k}=\bm{\theta}^{n,0}$; $Z^{n,1}=Z^{n-1}$\;
		\While{change $>$ tol}{
			$W^{\tm{h};n,k}=(1/B_\tm{r})\,\Delta x_1\, \Delta x_2 \sum_{i,h} p_{i,j}^{n,k-1}$\;
			$W^{\tm{c};n,k}=(1/B_\tm{r})\,\Delta x_1\, \Delta x_2 \sum_{i,h} p_{\tm{c};i,j}^{n,k-1}$\;
			$Z^{n,k}=Z^n+\Delta t\,V^{n-1}+\Delta t^2/(2\,m)(W^{\tm{h};n,k}+W^{\tm{c};n,k}+W^\tm{a}(t^n))$\;{compute $T(\bm{\theta}^{n,k-1})$}\;
			\For{$i=1\ldots N_{x_1}$, $j=1\ldots N_{x_2}$}{
				\If{$\left(C_{i,j}-e^{00}_{i,j}\right)/a^{00}_{i,j}\,\geq\, T(\bm{\theta}^{n,k-1})_{i,j}$}
				{$p_{i,j}^{n,k} = (C_{i,j}-e^{00}_{i,j})/a^{00}_{i,j}$\;
					$\theta_{i,j}^{n,k}=1$\;
					\Else{
						$\theta_{i,j}^{n,k}=(C_{i,j}-a^{00}_{i,j}\, T(\bm{\theta}^{n,k-1})_{i,j})/e^{00}_{i,j}$\;
						$p_{i,j}^{n,k} = T(\bm{\theta}^{n,k-1})_{i,j}$\;
					}
				}
			}
		    
			$change=\|\bm{p}^{n,k}-\bm{p}^{n,k-1}\|+\|\bm{\theta}^{n,k}-\bm{\theta}^{n,k-1}\|+\|Z^{n,k}-Z^{n,k-1}\|$\;
			$\bm{p}^{n,k+1}=\bm{p}^{n,k}$; $\bm{\theta}^{n,k+1}=\bm{\theta}^{n,k}$; $Z^{n,k+1}=Z^{n,k}$\;
			$k=k+1$\;
		}
		$V^n=V^{n-1}+(\Delta t/m) (W^{\tm{h};n,k}+W^{\tm{c};n,k}+W^\tm{a}(t^n))$\;
		return $\bm{p}^{n,k}$, $\bm{\theta}^{n,k}$, $Z^{n,k}$, $V^{n}$\;
	}
	\caption{Adaptation of the numerical algorithm presented by \cite{Alt1980} to solve system \eqref{eq:sec3:fixed_point}.}\label{alg:sec3:alg-gs_reynolds}
\end{algorithm}

\section{Numerical examples}
\label{sec:num-simulations}

The nonlinearity of the proposed model is quite strong. At each time, the rightmost connected component of the cavitated/separated region must be identified to compute the operator $T$. Further, as a way to impose the zero normal pressure gradient at rupture points, $T$ is extended by one cell outside the numerical boundary of this connected component. Though the model is consistent with the physics, it may well be ill-posed, which would result in a chaotic behavior of the numerical solution as the mesh and the time step are refined. Convergence analyses in non-trivial cases are thus crucial to assess the model and its numerical implementation. Such analyses, with increasing complexity, are reported in what follows.

For the following simulations the bore radius is set to $B_\tm{r}=41$ (4.2 cm), {the ring's length along $x_1$ to $L_\tm{ring}=1$ (1 mm)}, the ring mass to $m=1.25\times 10^{-5}$ per unit width ($0.03$ kg/m), the applied load to $W^\tm{a}=-1.666\times 10^{-4}$ per unit width ($-40$ N/m) {and the Back-pressure factor to $\gamma=0.9$}.

\subsection{One-dimensional stationary solutions}
\label{sec:num:stationary}

We begin by illustrating the model's behavior through a series of stationary solutions assuming different values for $\pcc$ (constant in time). The ring-to-liner distance $Z$ is also fixed to unity. The solutions are obtained by means of transient simulations for a time length $T$ big enough such that no variation on $\bm{p}$ nor $\bm{\theta}$ is observed for $t>T$. Based in a convergence analysis exposed in the next section, the cell size and time step length are chosen as $\Delta x_1=0.005$ and $\Delta t= 0.01$.

\begin{figure}[h!]
	\centering
	\def\svgwidth{\linewidth}
	\includegraphics[width=0.7\linewidth]{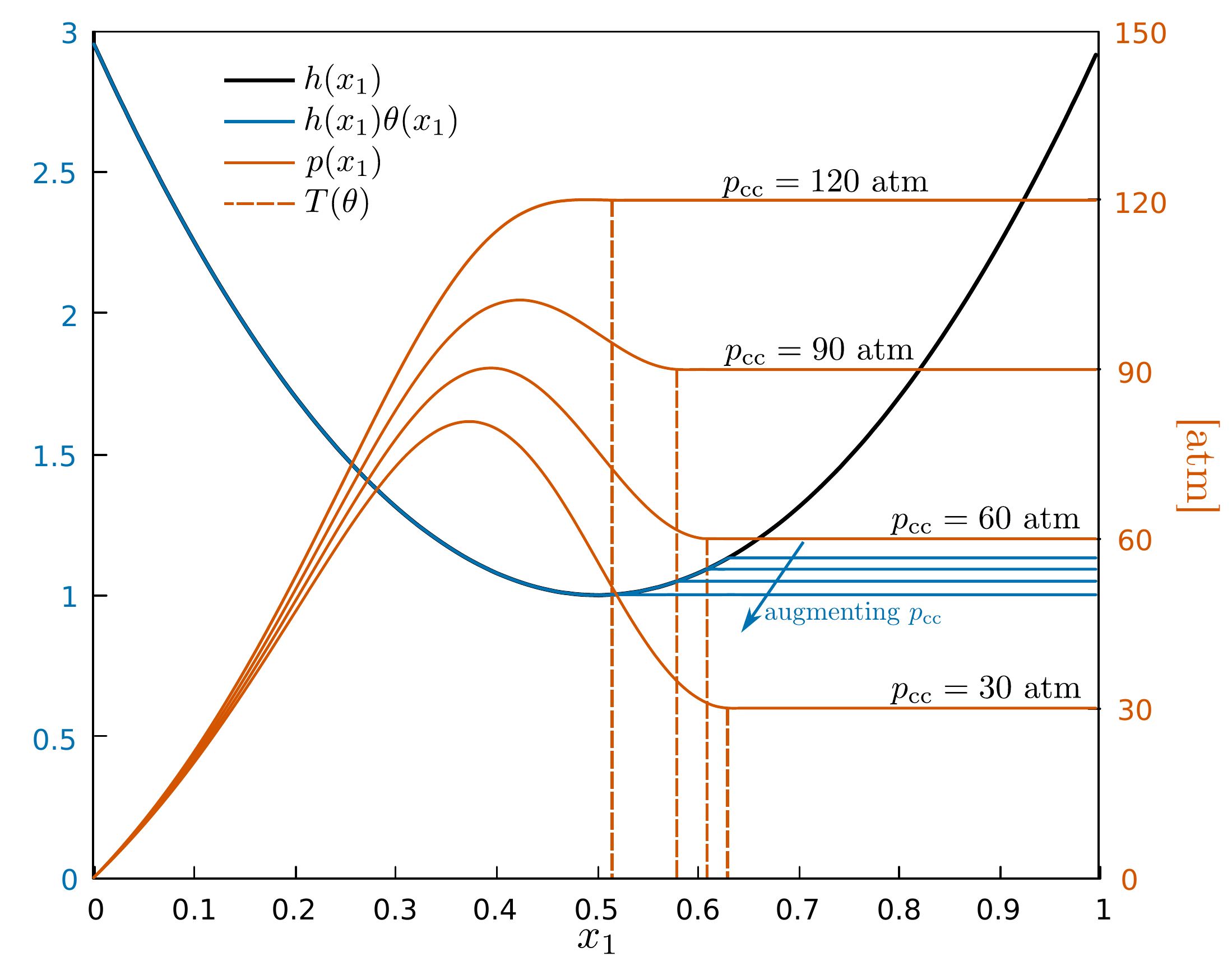}
	\caption{Stationary cases for $R=64$, $u=1$, $h_\tm{feed}=3$ and different values of $\pcc$.}
	\label{fig:sec4:stationary_cases}
\end{figure}

The pressure and saturation fields obtained for different values of $\pcc$ are shown in Fig. \ref{fig:sec4:stationary_cases} for the lower surface moving to the right ($u=1$), thus the point $\beta\in\partial \Omega_0^r$ corresponds to a rupture point. As indicated in that figure, the higher the value of $\pcc$ the lower the fluid film thickness that exits the domain. Moreover, there exists an upper limit on $\pcc$ above which the algorithm does not converge. This coincides with the value of $\pcc$ for which the rupture point reaches $x_1=0.5$, which is around $120$ atm. For higher values of $\pcc$ the simulation \emph{fails} in the sense that a maximum number of iterations (=100,000) is reached without the numerical variable ``\emph{change}'' becoming lower than ``\emph{tol}'' in Algorithm \ref{alg:sec3:alg-gs_reynolds}. Since stationary solutions with a rupture point placed at the left side of $x_1=0.5$ have no physical meaning, that value can be interpreted as the maximum CCP such that the ring is able to seal the combustion chamber and prevent blow-by. Based on these observations, the condition
\begin{equation}
\frac{\partial^2 p}{\partial x_1^2}\leq 0\qquad\mbox{and}\qquad p>0\label{eq:blow-by-cond-rupture}
\end{equation}
is identified as a blow-by prediction condition at rupture points.

\begin{figure}[h!]
	\centering
	\def\svgwidth{\linewidth}
	\includegraphics[width=0.7\linewidth]{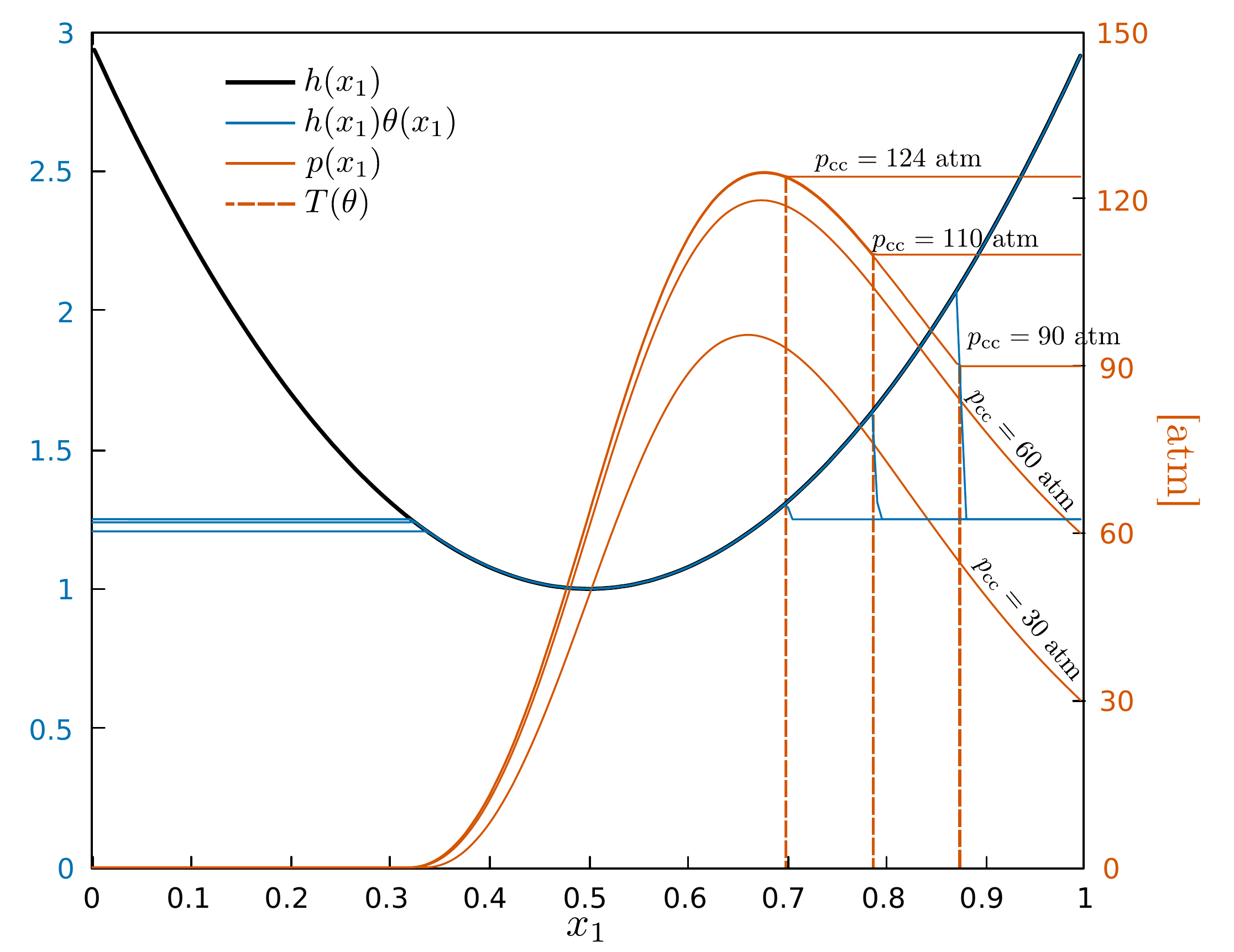}
	\caption{Stationary cases with $u=-1$, $h_\tm{feed}=1.25$ and different values of $\pcc$.}
	\label{fig:sec4:stationary_cases2}
\end{figure}

The pressure and saturation fields obtained for different values of $\pcc$ are shown in Fig. \ref{fig:sec4:stationary_cases2} for the lower surface moving to the left ($u=-1$) so that the cavitation boundary adjacent to the combustion chamber is a reformation point. Observe that for $\pcc>60$ atm the pressure profiles coincide to the left of $\beta$ (that depends on $\pcc$), only changing to the right of that point (where $p=\pcc$). For these cases the pressure gradient is not necessarily null (recall that $\theta_-$ can be lower than one in \eqref{eq:sec2:rankine}), thus, from the observed behavior the condition
\begin{equation}
\frac{\partial p}{\partial x_1}\geq 0\qquad\mbox{and}\qquad p>0\qquad\label{eq:blow-by-cond-reformation}
\end{equation}
is identified as a blow-by prediction condition at reformation points. 

\subsection{Rings with textures}
\label{sec:num:textured}
\begin{figure}[h!]
	\centering
	\def\svgwidth{0.9\linewidth}
	\includegraphics[width=1\linewidth]{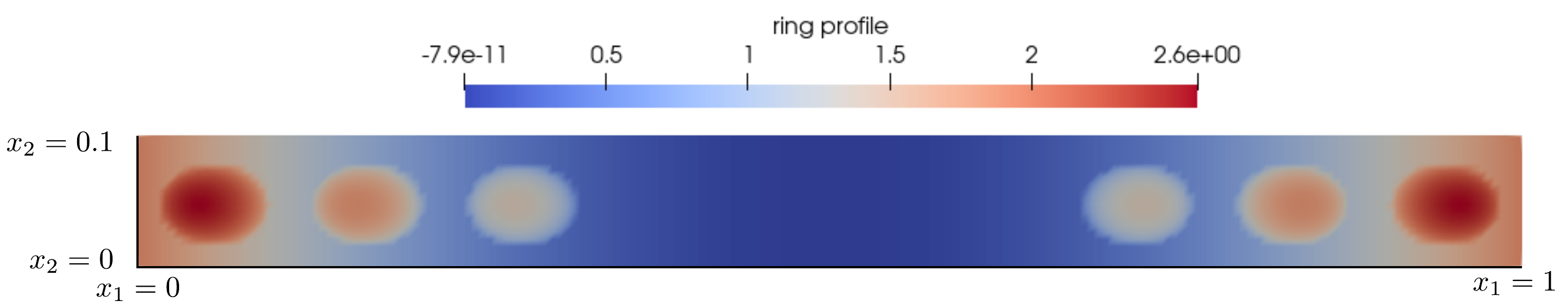}
	\caption{Texture set on the ring for the first reported simulations.}
	\label{fig:sec4:ring_texture}
\end{figure}

Consider now simulations with a texture consisting of elliptically shaped dimples on the ring. The dimples are arranged at a distance equal to 0.1 along the $x_2$ axis, its depth is set to 1 micron and their size along the $x_1$ and $x_2$ axis is fixed to 80 and 60 microns, respectively. The resulting ring's profile ($h_\tm{ring}(x_1,x_2)$) is detailed in Fig. \ref{fig:sec4:ring_texture}. This is a sufficiently complex case to numerically assess the well-posedness of the proposed model.

\subsubsection{Convergence in space}

The radial dynamics of the ring is solved as exposed in Algorithm \ref{alg:sec3:alg-gs_reynolds}, with an initial value $Z(t=0)=0.8$. The cell sizes are set by taking $\Delta x_1=\Delta x_2$ and by comparing stationary solutions obtained when fixing $\pcc$ to 50 atm. These stationary solutions consist of time-converged transient simulations for a period of 0.001 s, long enough for the time derivatives to be negligible. The time step is set such that the Courant-Friedrichs-Levy (CFL) number is equal to the unit.

\begin{figure}[h!]
	\centering
	\def\svgwidth{1\linewidth}
	\includegraphics[width=0.7\linewidth]{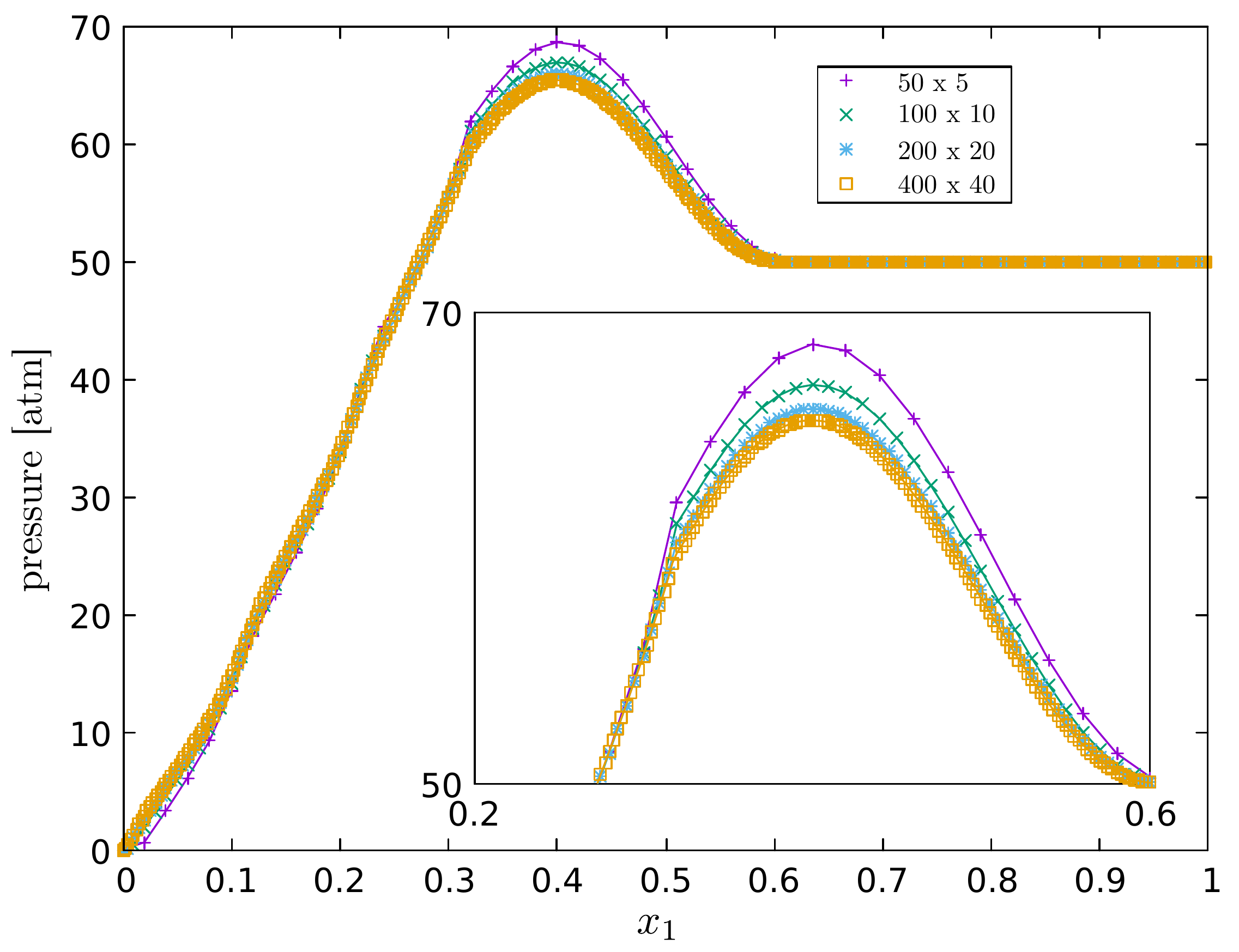}
	\caption{Convergence of $\bm{p}$ in the stationary regime {(with u=1)} along $x_2=0.05$ when varying the number of cells along $x_1$ and $x_2$.}
	\label{fig:sec4:convergence_p_mesh}
\end{figure}

Figure \ref{fig:sec4:convergence_p_mesh} shows pressure profiles obtained along the centerline of the computational domain (i.e., fixing $x_2=0.05$). Spatial convergence of the proposed algorithm is observed as the mesh is refined. Next, time convergence is addressed keeping the number of cells along the $x_1$-axis fixed to 200.

\subsubsection{Convergence in time and comparison with classical cavitation models}

\begin{figure}[h!]
	\centering
	\def\svgwidth{\linewidth}
	\includegraphics[width=0.7\linewidth]{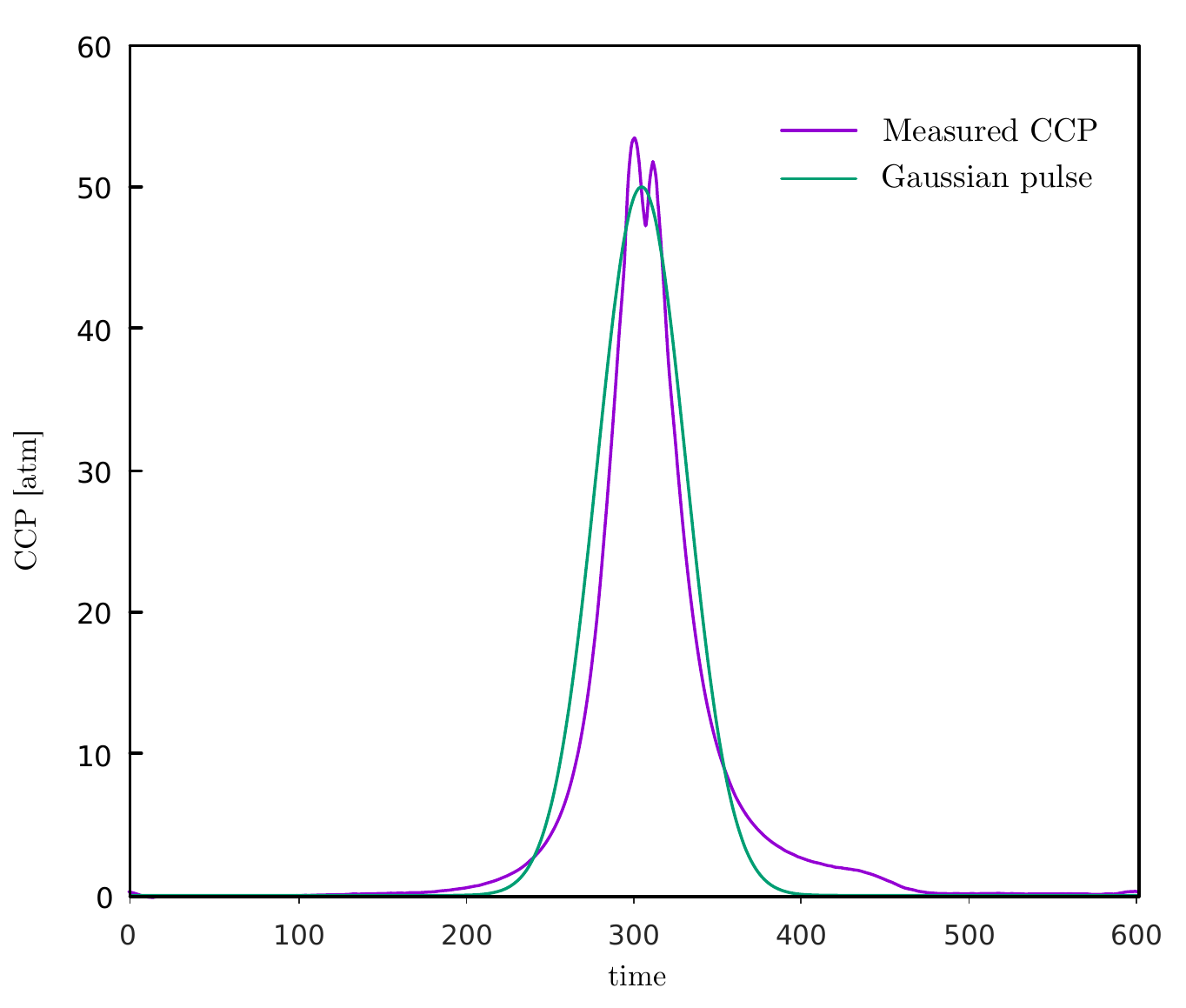}
	\caption{Combustion chamber pressure in time modeled as a Gaussian pulse.}
	\label{fig:sec4:pressure_time}
\end{figure}

In this transient simulations the CCP is modeled as a Gaussian pulse in time of duration $\simeq 0.02$ s with its amplitude denoted by $A_\tm{cc}$. In Fig. \ref{fig:sec4:pressure_time} it is shown that the Gaussian pulse indeed approximates a measured CCP (taken from \cite{Wakuri1992}) for $A_\tm{cc}=50$ atm. This example is quite challenging, not just because of the texture, but also because $\pcc(t)$ is non-monotonous, making $\Omega_0^r$ to grow and shrink dynamically.

\begin{figure}[h!]
	\centering
	\def\svgwidth{\linewidth}
	\includegraphics[width=0.7\linewidth]{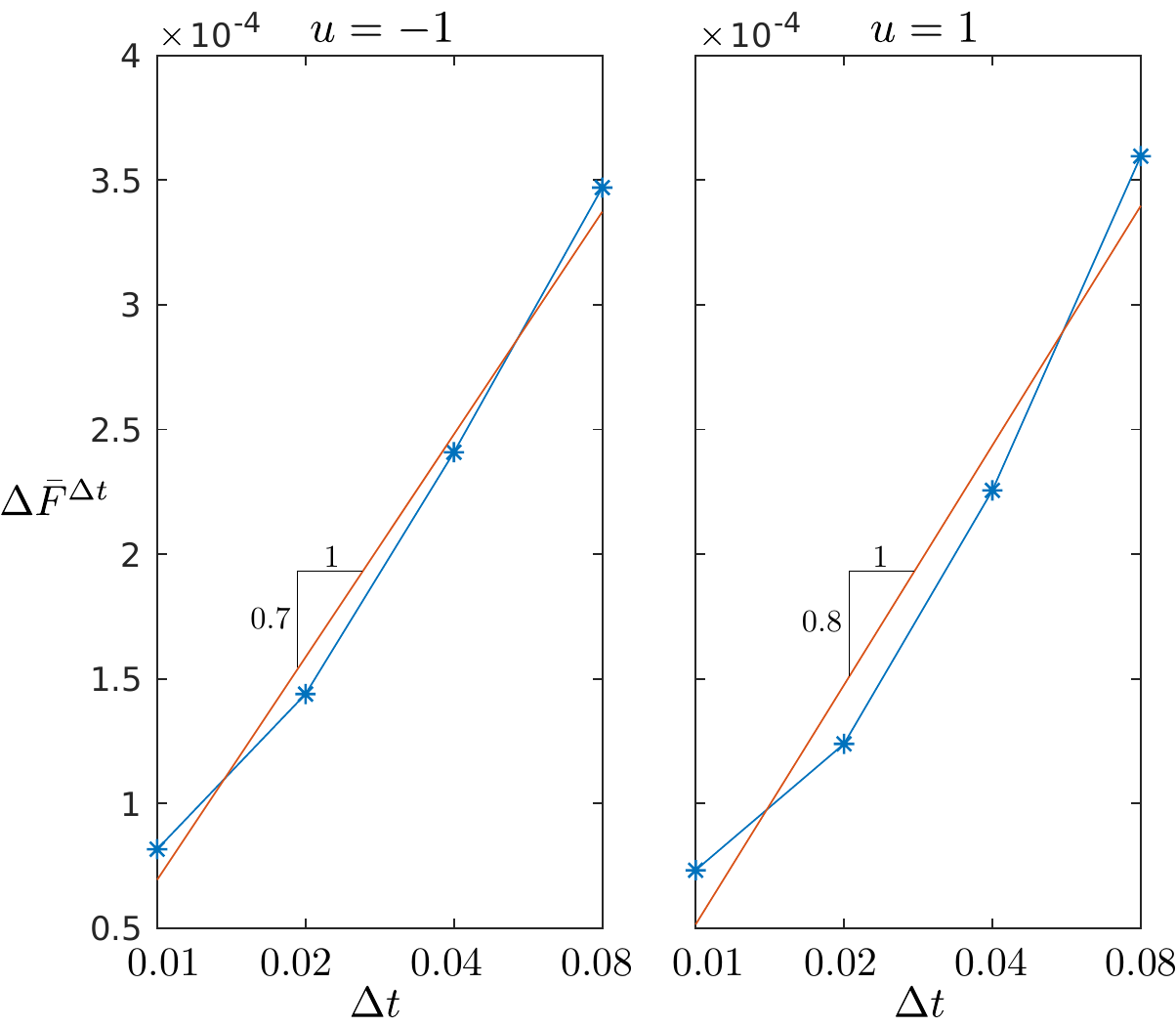}
	\caption{Relative difference in friction for different time steps $\Delta t$, $u=-1$ and $u=1$.}
	\label{fig:sec4:convergence_time}
\end{figure}

{A convergence analysis was performed by computing the relative difference on friction in time according to
$$\Delta \bar{F}^{\Delta t} = \frac{\|{F}^{\Delta t}(t)-{F}^{0.0025}(t)\|_1}{\|{F}^{0.0025}(t)\|_1}$$
where $F^{\Delta}$ is a discretization of \eqref{eq:sec2:friction} and $\|\cdot\|_1$ is a discretization of $\int_{0}^{600} |\cdot| \,dt$. Figure \ref{fig:sec4:convergence_time} shows $\Delta \bar{F}^{\Delta t}$ vs. $\Delta t$, considering both $u=-1$ (reformation boundary on the combustion-chamber side of the ring) and $u=1$ (rupture boundary). A convergence rate $\approx \Delta t^{0.7}$ is observed. An analogous analysis on the MFT dependence in $\Delta t$ showed a convergence rate $\approx \Delta t^{0.6}$. These results suggest} that the model is indeed well-posed and stably approximated, and it is thus interesting to compare it with the models previously available in the literature, that either consider $\pcc=0$ for the hydrodynamic problem (standard Elrod-Adams model) or are not mass-conserving (Reynolds cavitation model). For this comparison, the number of cells along $x_1$ is set to 200 and the time step is set to $2 \Delta x$.

\begin{figure}[h!]
	\centering
	\def\svgwidth{\linewidth}
	\includegraphics[width=0.8\linewidth]{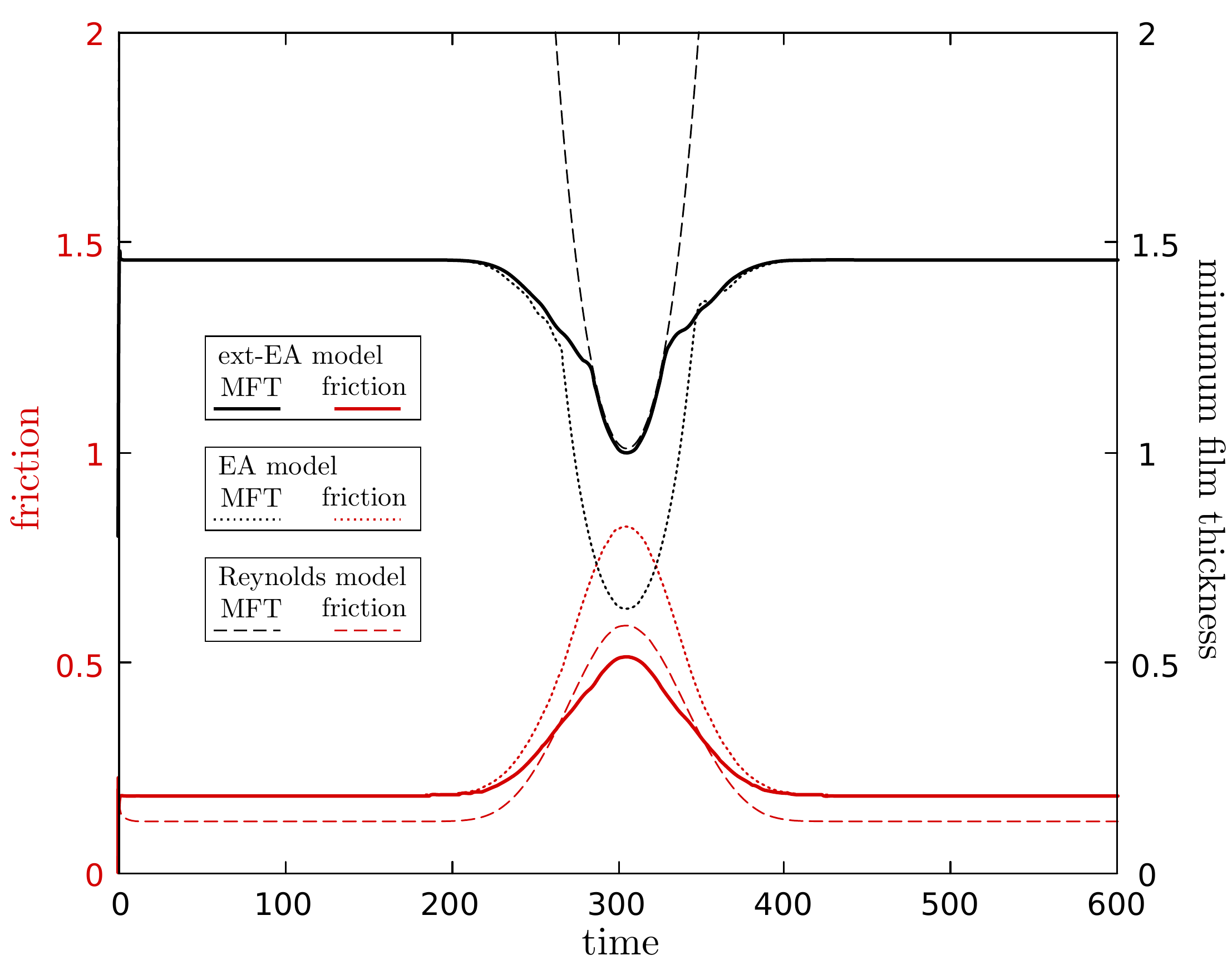}
	\caption{Friction force and MFT obtained with the extended (ext-EA), the Elrod-Adams (EA) and the Reynolds cavitation models.}
	\label{fig:sec4:friction_hmin_time}
\end{figure}

\begin{figure}[h!]
	\centering
	\def\svgwidth{\linewidth}
	\includegraphics[width=1\linewidth]{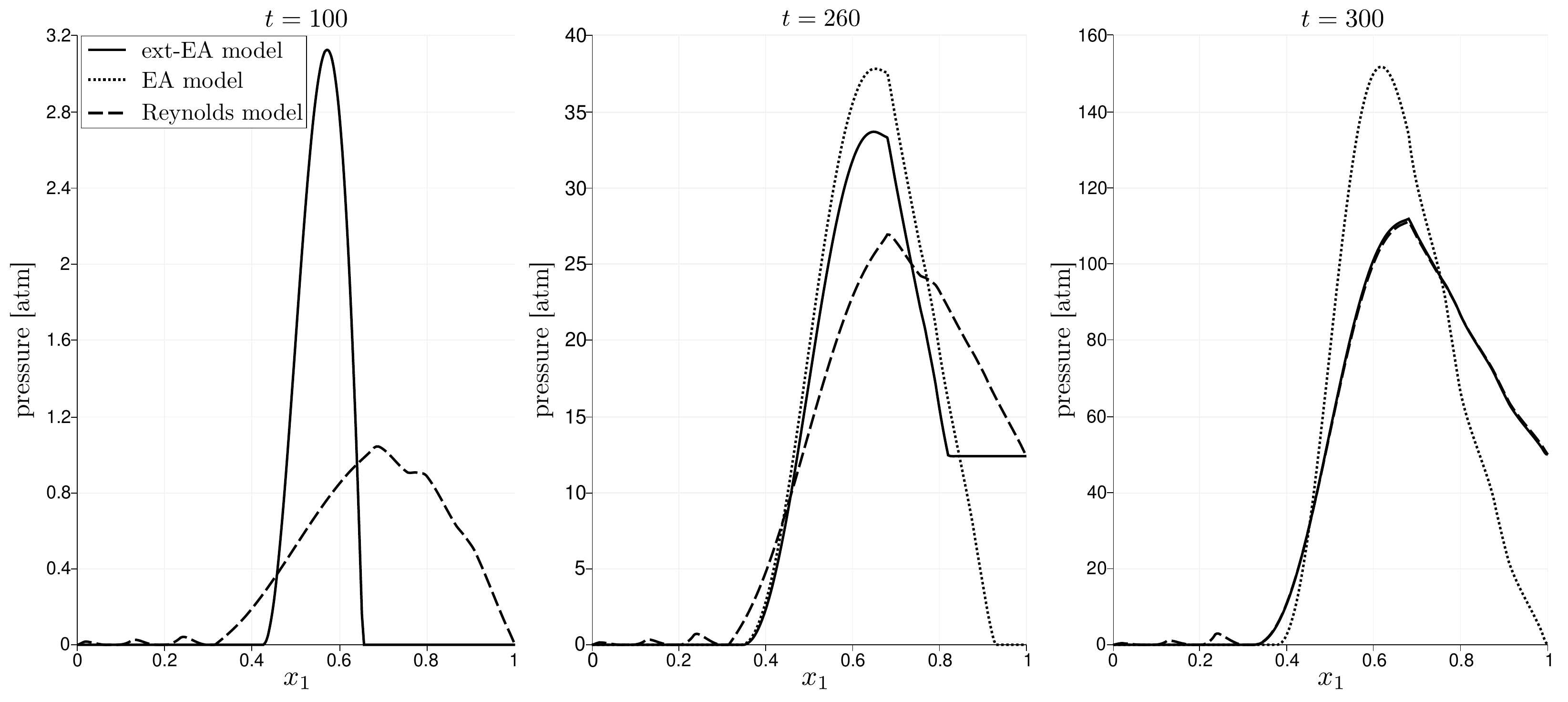}
	\caption{Pressure profiles fixing $x_2=\pi B_\tm{r}$ at $t=100$ (stationary regime, CCP $\simeq 0$), $t=260$ (CCP $\simeq 13$ atm), $t=300$ (CCP $\simeq 50$ atm) for different cavitation models.}
	\label{fig:sec4:comparison_pressure}
\end{figure}

The comparative results for friction and MFT are shown in Fig. \ref{fig:sec4:friction_hmin_time}. Notice that the simulated time ($0.06$ s) corresponds to a whole engine cycle running at 2000 rpm. To analyze these results the extended model is set as a reference. When comparing with the Elrod-Adams model, a difference of $\simeq 0.5$ microns is found for the MFT while the explosion is taking place, which translates into a 17\% of difference in the time-averaged friction. On the other hand, the Reynolds model predicts a much higher value of the MFT ($Z\simeq 6.8$) when $\pcc$ is negligible, while for $280<t<330$ the MFT is quite similar to the one predicted by the extended model. Overall, this implies a difference of 25\% in the time-averaged friction. The pressure profiles along $x_2=\pi B_\tm{r}$ are shown in Fig. \ref{fig:sec4:comparison_pressure} for different time instants. While the Elrod-Adams model is not able to impose $p=\pcc$ at $x_1=1$, the Reynolds model fails to predict the cavitation boundary correctly. Both these issues are resolved by the extended model, which provides a physically sound solution.

\noindent\textbf{Remark:} The friction formula used for the Reynolds model is similar to \eqref{eq:sec2:friction} but replacing the Couette term $\mu u / h\, g(\theta)$ by $$\begin{cases}
	\mu u / h & \mbox{in } \Omega_+\\
	0 & \mbox{in } \Omega_0
	\end{cases}.$$

\subsection{Simulation of the full engine cycle considering wear}
\label{sec:num:wear}

\begin{figure}[h!]
	\centering
	\def\svgwidth{\linewidth}
	\includegraphics[width=0.7\linewidth]{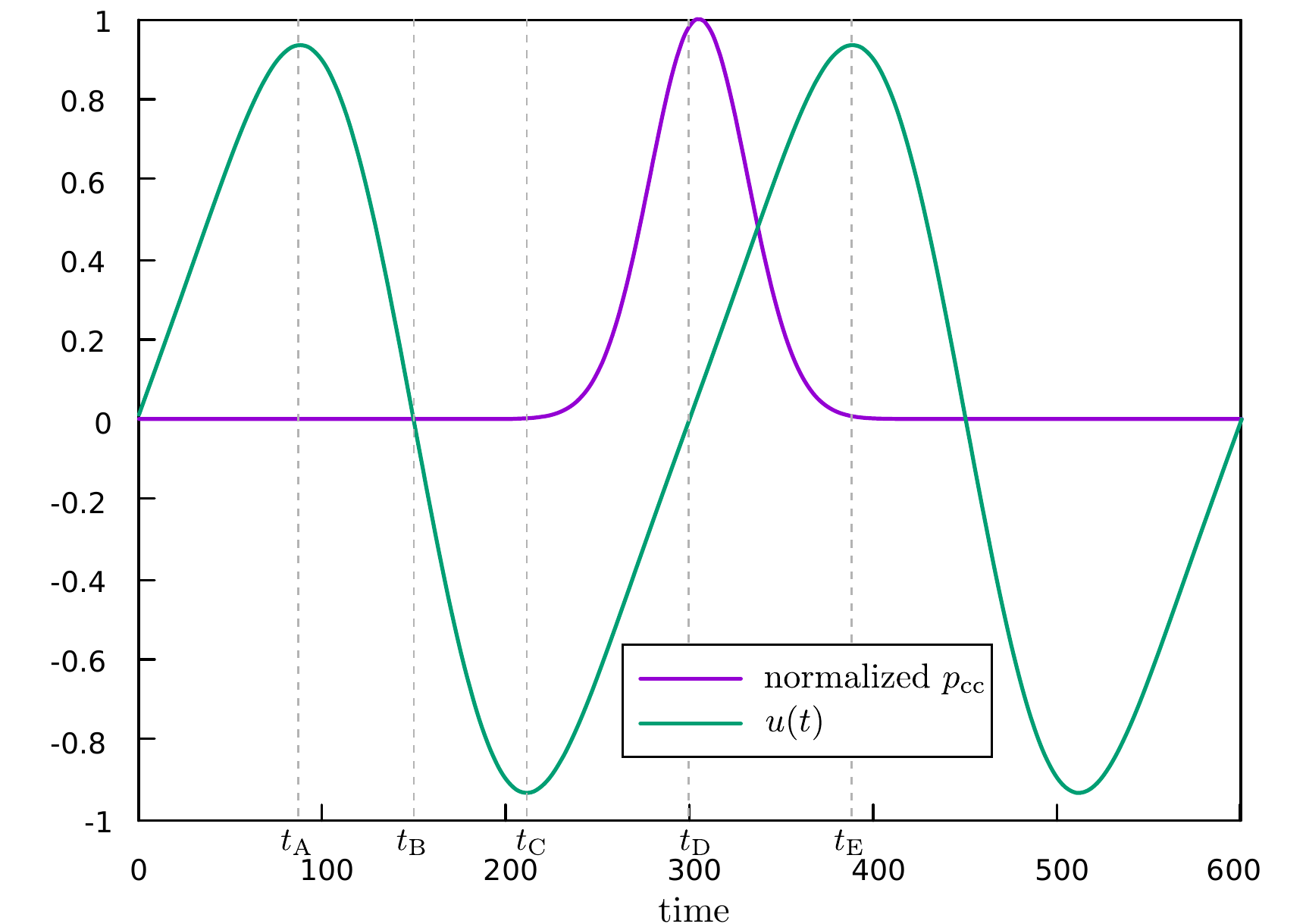}
	\caption{{Liner's} speed $u$ and CCP (normalized with its maximum value) as functions of time to model a full four-stroke engine cycle.}
	\label{fig:sec4:speed_pressure_engine}
\end{figure}

In this section both surfaces are assumed untextured, but a non-uniform wear is incorporated on the ring by taking $\delta\geq0$ in Eq. \eqref{eq:sec2:h-gap}. Notice that the wear can also be interpreted as taking place on the cylinder bore. Based on a convergence analysis analogous to the one presented in the previous section the cell size is set to $\Delta x_1=1/200$ and $\Delta x_2=2\pi B_\tm{r}/40$, while the Courant number is set to $1$ during most of the simulation. The film thickness at the fluid entrance is set to $h_\tm{feed}=1.5$. The liner's speed $u$ (relative to the ring) and the normalized variation of $\pcc(t)$ are shown in Fig. \ref{fig:sec4:speed_pressure_engine}. Some convergence difficulties that arise when $\pcc(t)$ is near its maximum and the speed of the ring is small (at the half of the cycle) were overcome by setting the Courant number to $0.05$ for $290<t<314$.

\subsubsection{Blow-by inception for different ring's curvatures}
\label{sec:blow-by-curvatures}
Here the height of the ring's profile at the boundaries is denoted by $\Delta h\overset{\scriptsize\mbox{def}}{=}h^R_\tm{ring}(x_1=1)$. A series of simulations are performed by varying $\Delta h$ as $0.5,1.0,1.5,2,\ldots, 8.0$ and $\delta$ as $0.005, 0.010, 0.015,\ldots,0.060$. From the results it is observed that for each $\Delta h$ there exists a value of $\delta$, denoted by $\delta^\tm{max}$, for which the simulation fails at a certain time step before the whole cycle is completed, and the same is observed for every $\delta\geq \delta^\tm{max}$, while for $\delta< \delta^\tm{max}$ the simulation is completed. These results on $\delta_\tm{max}$ are summarized in Table \ref{tab:hmax}. It is worth to notice that $\delta_\tm{max}$ is maximum (as a function of $\Delta h$) for $\Delta h$ in the interval $2.5-3.0$, or $R$ in $41.6-50$, which corresponds to typical values for the compression ring's curvature \cite{Gadeshi2012}. 

The time step at which the simulations fail varies with $\Delta h$ and $\delta$. In general, for $\Delta h \geq 3.91$ all the failures take place at $t\approx 300$, where $\pcc$ is maximum, while for $\Delta h < 3.91$ the time at which the simulations fail diminishes and the value of $\pcc$ at that instant is much lower. Next, we give more details on this, showing the pressure profiles around the failure time for different regimes.

\begin{table}[h!]\centering {\small
	\begin{tabular}{cccccccccc}
		\toprule
		\multirow{2}{*}{} &
		
		\multicolumn{7}{c}{$\Delta h${\scriptsize($R$)}} \\
		& 0.5{\scriptsize(250)} & 1.0{\scriptsize(125)} & 1.5{\scriptsize(83.3)} & 2.0{\scriptsize(62.5)} & 2.5{\scriptsize(50)} & 3.0{\scriptsize(41.7)} & 3.5{\scriptsize(35.7)} &3.91{\scriptsize(32.0)}& 4.0-8.0{\scriptsize(31.25-15.6)}\\
		\midrule
		$\delta^\tm{max}$ & 0.025 &0.035 & 0.040 & 0.045 & 0.050 & 0.055 & 0.025& 0.005&0\\
		\bottomrule
	\end{tabular}}
\caption{Minimum value of $\delta$ for which the simulations fail for every $\delta\geq \delta_\tm{max}$.}\label{tab:hmax}
\end{table}

\begin{figure}[h!]
	\centering
	\def\svgwidth{\linewidth}
	\includegraphics[width=0.9\linewidth]{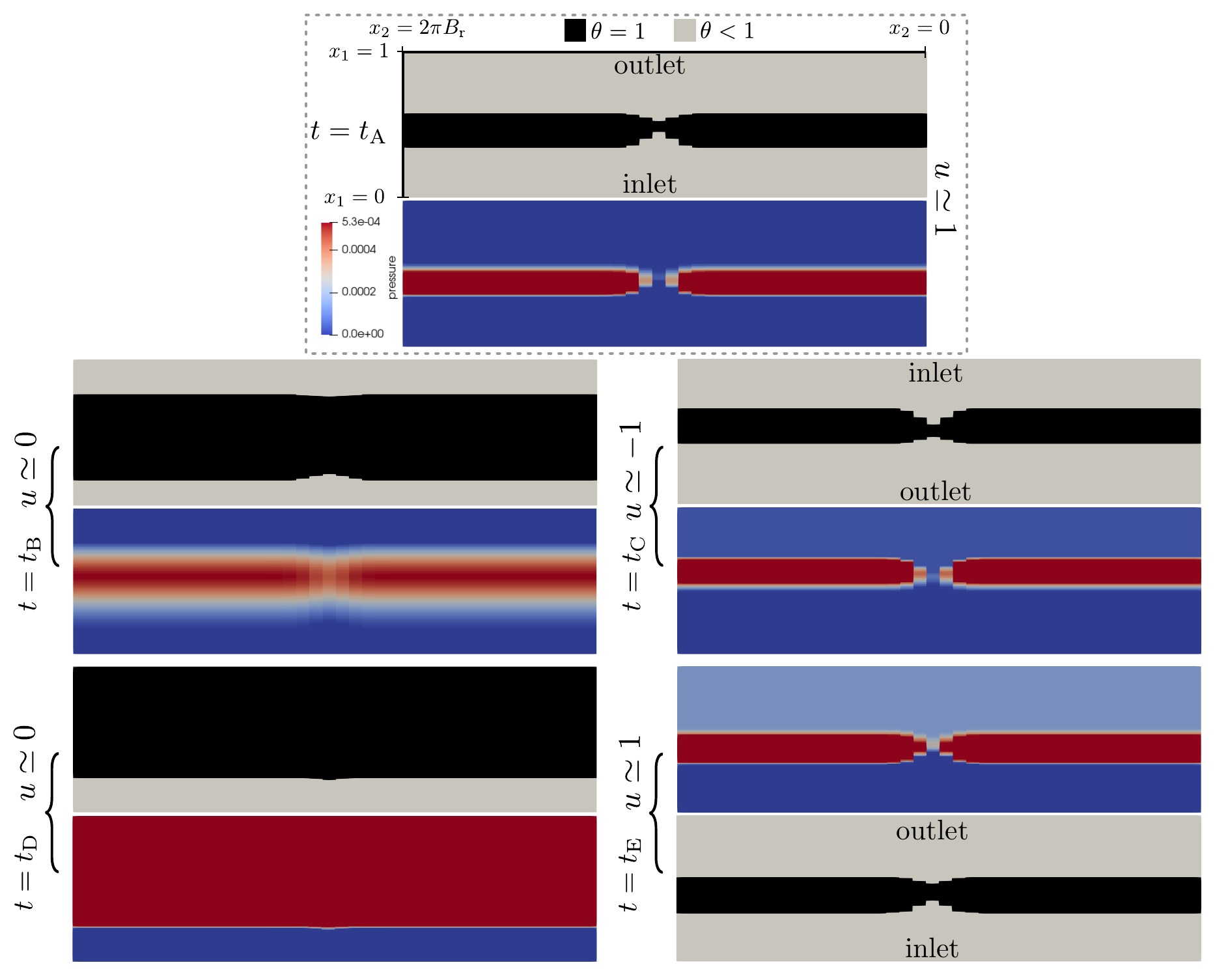}
	\caption{Fields $p$ and $\theta$ in time for a smooth ring with $R=62.5$ and $\delta=0.040$.}
	\label{fig:sec4:full-cycle-R_64-delta_004}
\end{figure}

Setting first $\Delta h =2$ ($R=62.5$) and the wear amplitude to $\delta=0.040$ one observes that the simulation is completed. Five time instants ($t_\tm{A}$, $t_\tm{B}$, ...) are chosen as indicated in Fig. \ref{fig:sec4:speed_pressure_engine} to show the pressure and saturation fields in Fig. \ref{fig:sec4:full-cycle-R_64-delta_004}. Notice that the full-film region (in black in the upper part of the figures for each time) separates two incomplete-film regions, one on the top that is connected to the combustion chamber, and one on the bottom where the pressure value is equal to 1 atm. The effect of the wear appears as a thinning of the full-film region around the circumferential position $x_2=\pi B_\textnormal{r}$, which corresponds to the sector of maximum wear. 

\begin{figure}[h!]
	\centering
	\def\svgwidth{\linewidth}
	\includegraphics[width=0.6\linewidth]{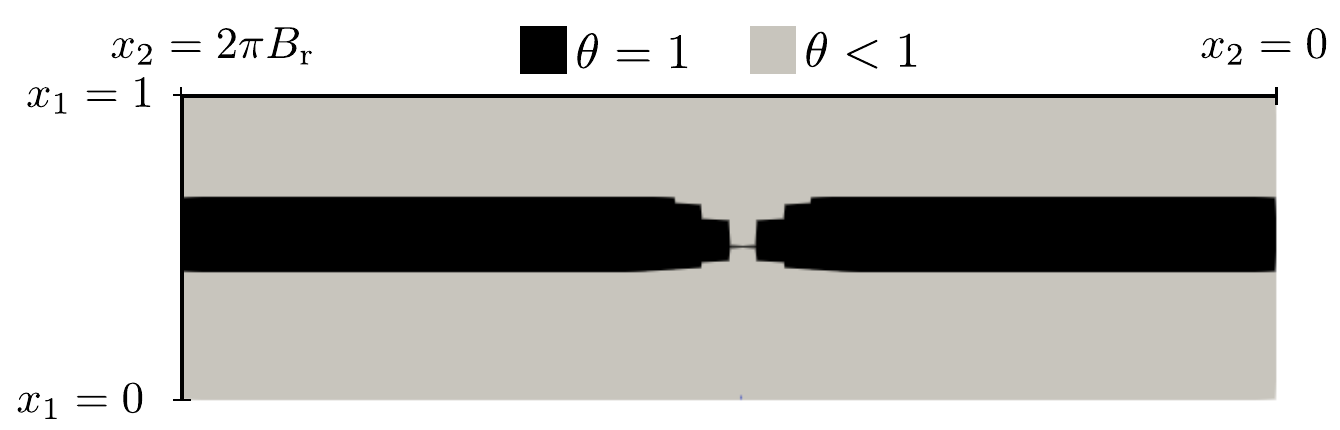}
	\caption{Field $\theta$ in at $t=203.18$ for $R=62.5$ and $\delta=0.045$.}
	\label{fig:sec4:theta_blow-by}
\end{figure}

\begin{figure}[h!]
	\centering
	\def\svgwidth{\linewidth}
	\includegraphics[width=0.7\linewidth]{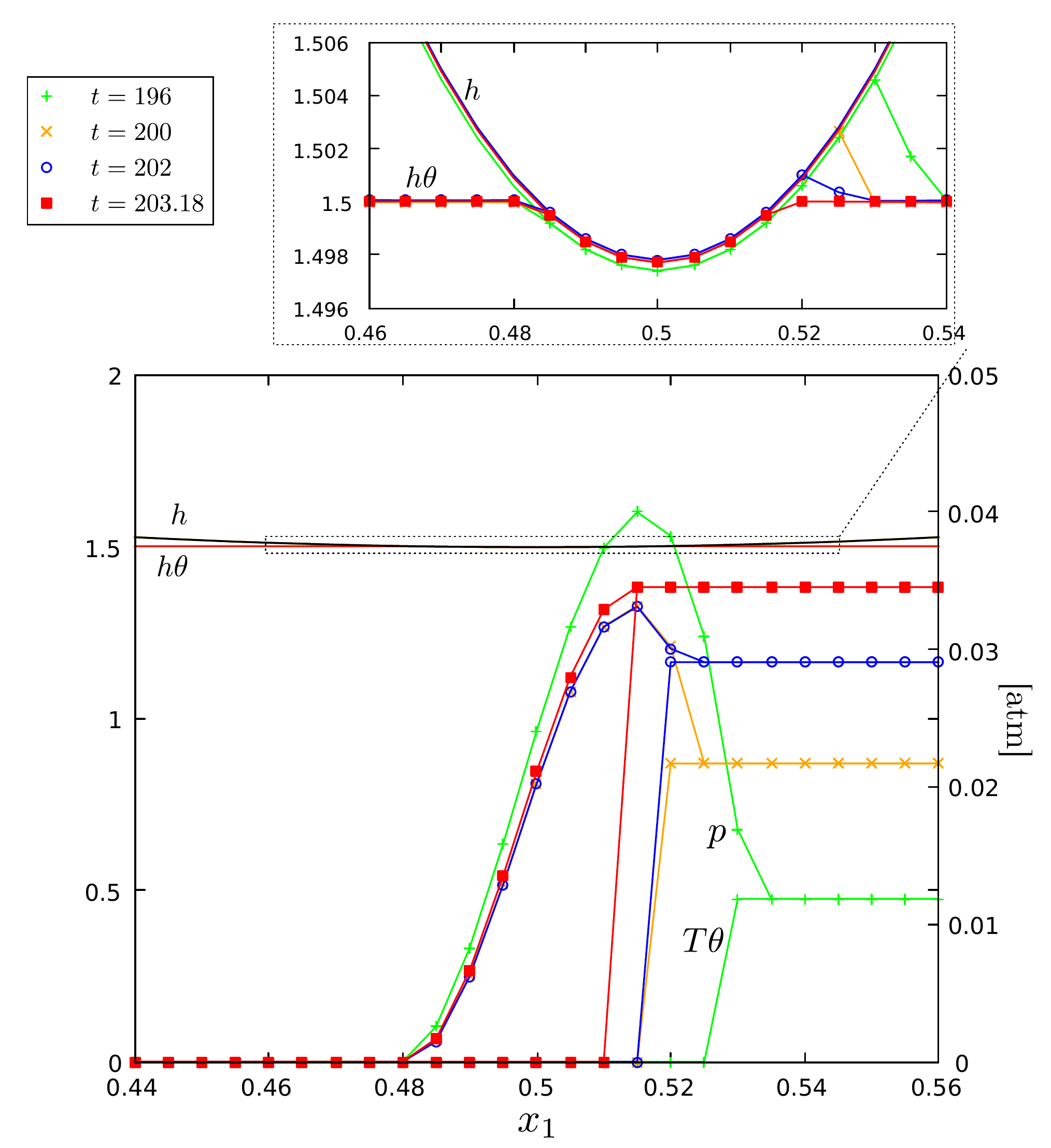}
	\caption{Pressure profiles and film thickness along $x_2=0.5$ for $\Delta h =2.0$ ($R=62.5$) and $\delta=\delta_\tm{max}=0.045$.}
	\label{fig:sec4:pres-blow-by-R62}
\end{figure}

Increasing $\delta$ to $0.045$ for the same ring curvature ($R=62.5$), one observes that as $t$ tends to $t=203$ the full-film region under the ring tends to break up, as shown in Fig. \ref{fig:sec4:theta_blow-by} . After $t=203.18$ the Algorithm \ref{alg:sec3:alg-gs_reynolds} fails to converge. At that simulation time, the CCP is equal to $0.035$ atm and our interpretation of the algorithm's failure is that the proposed model looses validity because blow-by takes place. Details
on the pressure profiles and film-thickness along $x_2=0.5$ are given in Fig. \ref{fig:sec4:pres-blow-by-R62}. Notice the drop of the pressure build-up in the full-film region $\{\theta=1\}$ as the value of $\pcc$ augments in time. 

\begin{figure}[h!]
	\centering
	\def\svgwidth{\linewidth}
	\includegraphics[width=0.7\linewidth]{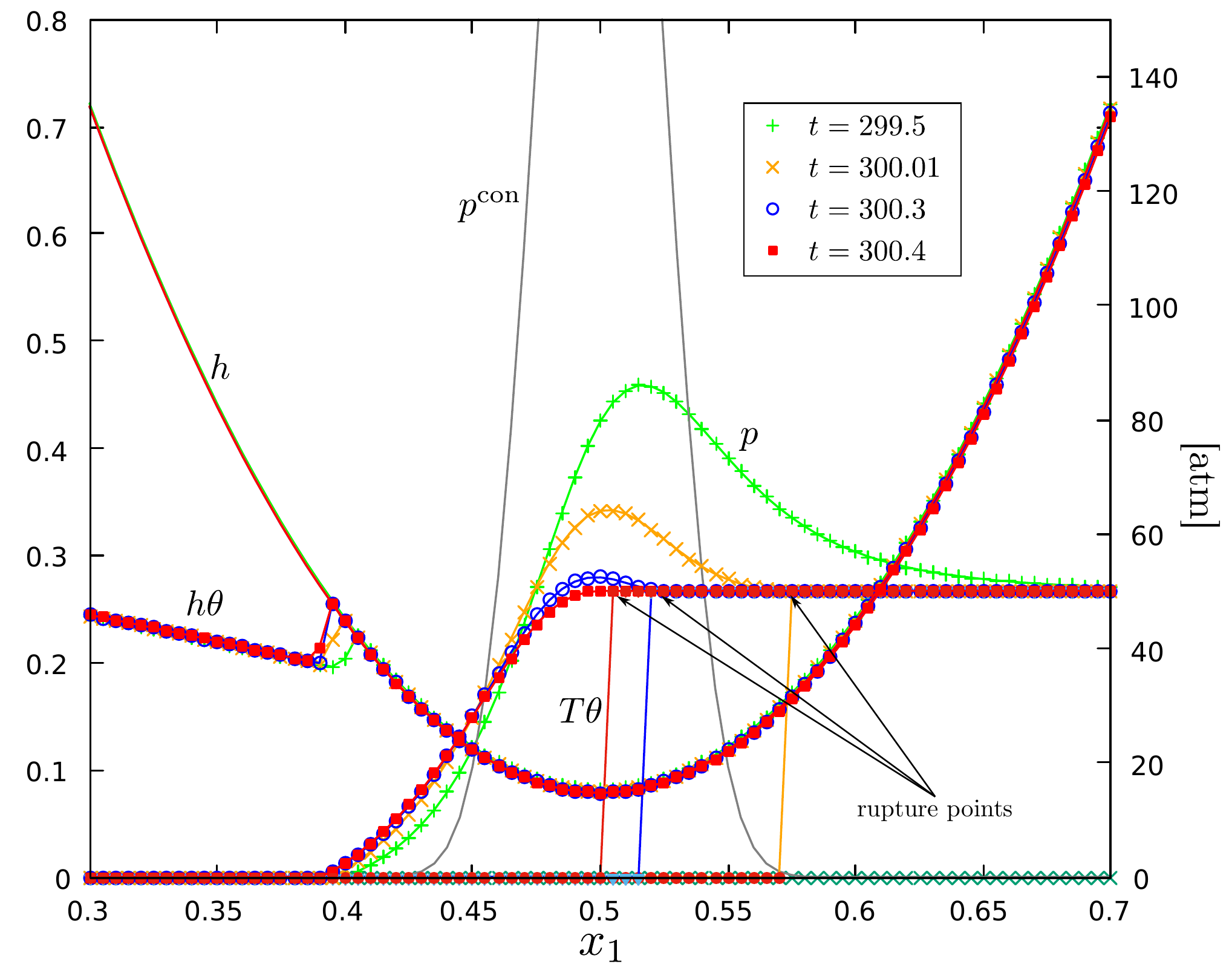}
	\caption{Pressure profiles and film thickness along $x_2=0.5$ for $\Delta h =4.0$ ($R=31.25$) and $\delta=\delta_\tm{max}=0$. The contact pressure is shown only for $t=299.5$ as it is quite similar for the other time steps.}
	\label{fig:sec4:pres-blow-by-R31}
\end{figure}

When $\Delta h$ is augmented to $3.91$ ($R=32$) the same phenomenon takes place even for $\delta$ as small as 0.005, while for $\Delta h =4$ ($R=31.25$) the simulations fails for every $\delta\geq 0$ tested. For this last case some of the pressure profiles and film-thickness along $x_2=0.5$ in time are detailed in Fig. \ref{fig:sec4:pres-blow-by-R31}. As for the previous case, the pressure build-up in the full-film region decreases and the rupture points (for $t>300$) advance to the left.

\subsubsection{A blow-by inception criterion}

\begin{figure}[h!]
	\centering
	\def\svgwidth{\linewidth}
	\includegraphics[width=0.7\linewidth]{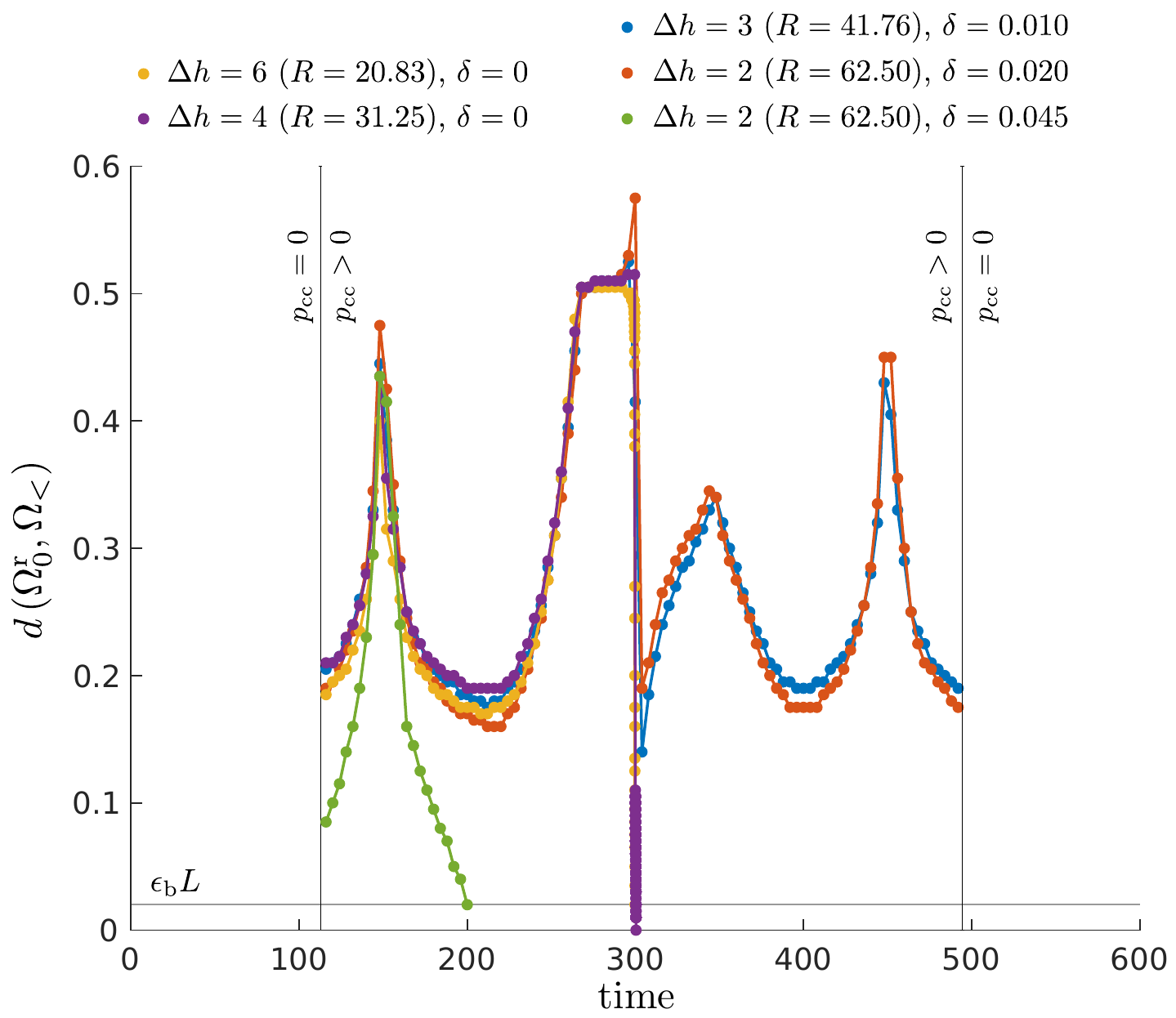}
	\caption{Distance between the sets $\Omega_0^\tm{r}$ and $\Omega_<$ for different configurations where. Blow-by is observed when $d\left(\Omega_0^\tm{r},\Omega_<\right)<\epsilon_\tm{b}L$.}
	\label{fig:sec4:blow-by-cond}
\end{figure}

In this section we focus on simulations that ended because of lack of convergence of Algorithm \ref{alg:sec3:alg-gs_reynolds} (defining a maximum number of iterations equal to $100,000$). We show that this lack of convergence can be related to blow-by inception as identified in the previous section. Further, and encompassing condition \eqref{eq:blow-by-cond-rupture} (for rupture boundary) and condition \eqref{eq:blow-by-cond-reformation} (for reformation boundary) which are hard to test numerically, we propose the condition
\begin{equation}
d\left(\Omega_0^r,\Omega_<\right)\leq \max\{\epsilon_\tm{b}L,N_\tm{b}\Delta x_1\} \qquad\mbox{and}\qquad\pcc>0,\label{eq:discrete-blow-by-condition}
\end{equation}
where $\Omega_<$ is the region of the domain where $p<\pcc$, while $d\left(\Omega_0^\tm{r},\Omega_<\right)$ denotes the distance between the set $\Omega_0^\tm{r}$ (where $p=\pcc$) and $\Omega_<$. Essentially, this condition expresses that if the full-film region separating the gas region in which $p=\pcc$ from the region where $p<\pcc$ is narrow enough, blow-by will happen. 
In this work we take $\epsilon_\tm{b}=1/50=0.02$ and $N_\tm{b}=4$, but the results are not much sensitive to this specific choice. In fact, a convergence analysis showed that  $d\left(\Omega_0^\tm{r},\Omega_<\right)$, as a function of time, becomes mesh-independent once the mesh is fine enough. Figure \ref{fig:sec4:blow-by-cond} shows $d\left(\Omega_0^\tm{r},\Omega_<\right)$ for different configurations. For the ones where the simulation failed, it is observed that the condition \eqref{eq:discrete-blow-by-condition} is reached some time steps before the end of the simulation. While for the case where the whole cycle is completed the condition \eqref{eq:discrete-blow-by-condition} is never reached. This qualifies \eqref{eq:discrete-blow-by-condition} as a criterion to numerically predict blow-by.

\section{Conclusions}
	\label{sec:conclusions}
	
An extension of the Elrod-Adams model that accommodates non-homogeneous boundary conditions for pressure has been presented and it has been applied to the PRL system. A mass-conserving algorithm that imposes a non-negative pressure gradient condition at the rupture boundary is used, which exhibits convergence in both time and space. To our knowledge, these are the first mass-conserving transient simulations for the PRL mechanism that predict a rupture boundary with a non-zero value of $\pcc$ enforced at it.

Full-engine cycle simulations were also presented, which showed significant differences in the friction computed by means of the extended model when compared to the one obtained with either the standard Elrod-Adams or the Reynolds models, none of which is able to correctly predict the position of the pressurized regions.
	
A perturbation on the ring's surface along the circumferential direction was assumed to model a non-uniform ring wear. For a big-enough perturbation the separated region where the pressure value is equal to the CCP (or $\pcc(t)$) reaches the separated region where the pressure is zero, which is interpreted as leading to blow-by. Based on observations of stationary pressure profiles, two analytic conditions on the derivatives of $p$ at the rupture/reformation points were proposed to predict blow-by inception. Furthermore, an  equivalent and easily implementable numerical criterion (observed to be mesh-independent) was proposed, and it was shown that users can rely on it to predict blow-by from numerical simulations of PRL systems. 
	
\section*{Acknowledgments}

The authors thank the financial support provided by CAPES (grant PROEX-8434433/D), FAPESP (grants 2013/07375-0 and 2018/08752-5) and CNPq (grant 305599/2017-8).

\bibliographystyle{unsrt}

\begin{thebibliography}{10}
	
	\bibitem{Dowson1979}
	D.~Dowson and C.~Taylor.
	\newblock {Cavitation in bearings}.
	\newblock {\em {Annu. Rev. Fluid Mech.}}, 11:35--66, 1979.
	
	\bibitem{Jeng1992}
	Yeau-Ren Jeng.
	\newblock {Theoretical Analysis of Piston-Ring Lubrication Part {I} - Fully
		Flooded Lubrication}.
	\newblock {\em {Tribol. T.}}, 35(4):696--706, 1992.
	
	\bibitem{Han1998}
	D.~C. Han and J.~S. Lee.
	\newblock {Analysis of the piston ring lubrication with a new boundary
		condition}.
	\newblock {\em {Tribol. Int.}}, 31(12):753--760, 1998.
	
	\bibitem{Mufti2008}
	R.~A.~a Mufti, M.~Priest, and R.~J. Chittenden.
	\newblock {Analysis of piston assembly friction using the indicated mean
		effective pressure experimental method to validate mathematical models}.
	\newblock {\em Proceedings of the Institution of Mechanical Engineers, Part D:
		Journal of Automobile Engineering}, 222(8):1441--1457, 2008.
	
	\bibitem{Gadeshi2012}
	G.~B. Gadeschi, K.~Backhaus, and G.~Knoll.
	\newblock {Numerical Analysis of Laser-Textured Piston-Rings in the
		Hydrodynamic Lubrication Regime}.
	\newblock {\em {ASME J. Tribol.}}, 134(4):1--8, October 2012.
	
	\bibitem{Mezghani2012}
	S.~Mezghani, I.~Demirci, H.~Zahouani, and M.~El Mansori.
	\newblock The effect of groove texture patterns on piston-ring pack friction.
	\newblock {\em {Precision Engineering}}, 36(2):210 -- 217, 2012.
	
	\bibitem{Meng2014}
	F.~Meng, X.~Wang, T.~Li, and Y.~Chen.
	\newblock {Influence of cylinder liner vibration on lateral motion and
		tribological behaviors for piston in internal combustion engine}.
	\newblock {\em {Proc. Inst. Mech. Eng. Part J J. Eng. Tribol.}}, 2014.
	
	\bibitem{Morris2014}
	N.~Morris, M.~Leighton, M.~{De la Cruz}, R.~Rahmani, H.~Rahnejat, and
	S.~Howell-Smith.
	\newblock {Combined numerical and experimental investigation of the
		micro-hydrodynamics of chevron-based textured patterns influencing
		conjunctional friction of sliding contacts}.
	\newblock {\em Proc. Inst. Mech. Eng. Part J J. Eng. Tribol.}, (4), nov 2014.
	
	\bibitem{Medina2015}
	S.~Medina, M.~T. Fowell, S.-C. Vladescu, T.~Reddyhoff, I.~Pegg, A.~V. Olver,
	and D.~Dini.
	\newblock {Transient effects in lubricated textured bearings}.
	\newblock {\em {Proc IMech, Part J: J Engineering Tribology}}, 229(4):523--537,
	2015.
	
	\bibitem{Kligerman2015}
	Y.~Kligerman and A.~Shinkarenko.
	\newblock {Analysis of friction in surface textured components of reciprocating
		mechanism}.
	\newblock {\em Proc. Inst. Mech. Eng. Part J J. Eng. Tribol.}, 229(4):336--349,
	apr 2015.
	
	\bibitem{Liu2016}
	W.~Liu, Z.~Huang, Q.~Liu, and J.~Zeng.
	\newblock {An isogeometric analysis approach for solving the Reynolds equation
		in lubricated piston dynamics}.
	\newblock {\em {Tribol. Int.}}, 103:149--166, 2016.
	
	\bibitem{Usman2016}
	A.~Usman and C.~W. Park.
	\newblock {Optimizing the tribological performance of textured piston
		ring-liner contact for reduced frictional losses in SI engine: Warm operating
		conditions}.
	\newblock {\em {Tribol. Int.}}, 99:224--236, 2016.
	
	\bibitem{Fang2017}
	C.~Fang, X.~Meng, and Y.~Xie.
	\newblock {A piston tribodynamic model with deterministic consideration of
		skirt surface grooves}.
	\newblock {\em {Tribol. Int.}}, 110:232--251, 2017.
	
	\bibitem{Ausas2007}
	R.~Ausas, P.~Ragot, J.~Leiva, M.~Jai, G.~Bayada, and G.~Buscaglia.
	\newblock {The impact of the Cavitation model in the Analysis of Micro-Textured
		Lubricated Journal bearings.}
	\newblock {\em {ASME J. Tribol.}}, 129(4):868--875, 2007.
	
	\bibitem{Qiu2009}
	Y.~Qiu and M.~M. Khonsari.
	\newblock {On the Prediction of Cavitation in Dimples Using a Mass-Conservative
		Algorithm}.
	\newblock {\em {ASME J. Tribol.}}, 131(4):041702--1, 2009.
	
	\bibitem{Chong2011}
	W.~W.~F. Chong, M.~Teodorescu, and N.~D. Vaughan.
	\newblock {Cavitation induced starvation for piston-ring/liner tribological
		conjunction}.
	\newblock {\em {Tribol. Int.}}, 44(4):483--497, 2011.
	
	\bibitem{Tomanik2013}
	E.~Tomanik.
	\newblock {Modelling the hydrodynamic support of cylinder bore and piston rings
		with laser textured surfaces}.
	\newblock {\em {Tribol. Int.}}, 59:90--96, 2013.
	
	\bibitem{Checo2014a}
	H.~M. Checo, R.~Ausas, M.~Jai, J.~Cadalen, F.~Choukroun, and G.~C. Buscaglia.
	\newblock {Moving textures: Simulation of a ring sliding on a textured liner}.
	\newblock {\em {Tribol. Int.}}, 72:131--142, 2014.
	
	\bibitem{Checo2014b}
	H.~M. Checo, A.~Jaramillo, M.~Jai, and G.~C. Buscaglia.
	\newblock {Texture-Induced cavitation bubbles and friction reduction in the
		Elrod-Adams Model}.
	\newblock {\em {Proc IMech, Part J: J Engineering Tribology}}, 229(4):478--492,
	2015.
	
	\bibitem{Profito2016}
	F.~J. Profito, E.~Tomanik, and D.~C. Zachariadis.
	\newblock {Effect of cylinder liner wear on the mixed lubrication regime of
		TLOCRs}.
	\newblock {\em {Tribol. Int.}}, 93:723--732, 2016.
	
	\bibitem{Profito2017}
	F.~J. Profito, S.~Vlădescu, T.~Reddyhoff, and D.~Dini.
	\newblock {Transient experimental and modelling studies of laser-textured
		micro-grooved surfaces with a focus on piston-ring cylinder liner contacts}.
	\newblock {\em {Tribol. Int.}}, 113:125--136, 2017.
	
	\bibitem{Checo2017}
	H.~Checo, A.~Jaramillo, R.~Ausas, M.~Jai, and G.~Buscaglia.
	\newblock {Down to the roughness scale assessment of piston-ring/liner
		contacts}.
	\newblock {\em IOP Conference Series: Materials Science and Engineering},
	174(1):012035, 2017.
	
	\bibitem{Hu2018}
	Y.~Hu, X.~Meng, and Y.~Xie.
	\newblock {A new efficient flow continuity lubrication model for the piston
		ring-pack with consideration of oil storage of the cross-hatched texture}.
	\newblock {\em {Tribol. Int.}}, 119:443--463, 2018.
	
	\bibitem{Jaramillo2016a}
	A.~Jaramillo, H.~Checo, and G.~Buscaglia.
	\newblock {Incorporation of back-pressure effects in the modeling of the
		cylinder/piston-rings system}.
	\newblock {\em {Proceedings of the XXXVII Iberian Latin American Congress on
			Computational Methods in Engineering - CILAMCE2016}}, 2016.
	
	\bibitem{Jaramillo2016b}
	A.~Jaramillo, H.~M. Checo, and G.~Buscaglia.
	\newblock {Non-homogeneous boundary conditions and cavitation modeling for
		Reynolds equation}.
	\newblock {\em {Mec\'anica Computacional}}, XXXIX(4):2087--2100, 2016.
	
	\bibitem{Cheng2015}
	J.~Cheng, X.~Meng, Y.~Xie, and W.~Li.
	\newblock {On the running-in behavior of rough surface of piston rings in mixed
		lubrication regime}.
	\newblock {\em Industrial Lubrication and Tribology}, 67(5):468--485, 2015.
	
	\bibitem{Greenwood1970}
	J.~A. Greenwood and J.~H. Tripp.
	\newblock {The Contact of Two Nominally Flat Rough Surfaces}.
	\newblock {\em {Proceedings of the {I}nstitution of {M}echanical {E}ngineers}},
	185(1):625--633, 1970.
	
	\bibitem{Panayi2008}
	A.~Panayi and H.~Schock.
	\newblock {Approximation of the integral of the asperity height distribution
		for the Greenwood--Tripp asperity contact model}.
	\newblock {\em {Proc IMech, Part J: J Engineering Tribology}}, 222:165--169,
	2008.
	
	\bibitem{Checophd}
	H.~M. Checo.
	\newblock {\em Models and methods for the direct simulation of rough and
		micropatterned surfaces}.
	\newblock PhD thesis, Instituto de Ci\^encias Matem\'aticas e de
	Computa\c{c}\~ao, USP, 2015.
	
	\bibitem{Jaramillophd}
	Jaramillo A.
	\newblock {\em New models and Numerical Methods in Hydrodynamic Lubrication}.
	\newblock PhD thesis, Instituto de Ci\^encias Matem\'aticas e de
	Computa\c{c}\~ao, USP, to be published.
	
	\bibitem{Buscaglia2013}
	G.~C. Buscaglia, I.~Ciuperca, E.~Dalissier, and M.~Jai.
	\newblock A new cavitation model in lubrication: the case of two-zone
	cavitation.
	\newblock {\em {Journal of Engineering Mathematics}}, 83:57--79, 2013.
	
	\bibitem{Ausas2013}
	R.~Ausas, M.~Jai, I.~S. Ciuperca, and G.~C. Buscaglia.
	\newblock {Conservative one-dimensional finite volume discretization of a new
		cavitation model for piston-ring lubrication}.
	\newblock {\em {Tribol. Int.}}, 57:54--66, 2013.
	
	\bibitem{Alt1980}
	H.~W. Alt.
	\newblock {Numerical solution of steady-state porous flow free boundary
		problems}.
	\newblock {\em {Numer. Math.}}, 36(1):73--98, 1980.
	
	\bibitem{Marini1986}
	L.~D. Marini and P.~Pietra.
	\newblock Fixed-point algorithms for stationary flow in porous media.
	\newblock {\em {Comput. Methods Appl. Mech. Eng.}}, 56(1):17--45, May 1986.
	
	\bibitem{Ausas2009}
	R.~Ausas, M.~Jai, and G.~Buscaglia.
	\newblock {A Mass-Conserving Algorithm for Dynamical Lubrication Problems With
		Cavitation}.
	\newblock {\em {ASME J. Tribol.}}, 131, 2009.
	
	\bibitem{Wakuri1992}
	Y.~Wakuri, T.~Hamatake, M.~Soejima, and T.~Kitahara.
	\newblock {Piston ring friction in internal combustion engines}.
	\newblock {\em Tribol. Int.}, 25(5):299--308, jan 1992.
	
\end{thebibliography}

\end{document}